\documentclass{elsart}

\usepackage[utf8]{inputenc} 
\usepackage{wasysym}

\usepackage{amssymb}

\usepackage{graphicx}

\begin{document}
\begin{frontmatter}

\title
{Observational evidence of the accelerated expansion of the universe}

\author{Pierre Astier and Reynald Pain}

\address{Laboratoire de Physique Nucl\'eaire et de Hautes Energies, 
Universit\'e Pierre et Marie Curie, Universit\'e Paris Diderot and CNRS/IN2P3,
4 place Jussieu, 75005 Paris, France}
\begin{abstract}
The discovery of cosmic acceleration is one of the most important
developments in modern cosmology. The observation,
thirteen years ago, that type Ia supernovae appear dimmer that they
would have been in a decelerating universe followed by a series of
independent observations involving galaxies and cluster of galaxies as
well as the cosmic microwave background, all point in the same
direction: we seem to be living in a flat universe whose expansion is
currently undergoing an acceleration phase. In this paper, we review
the various observational evidences, most of them gathered in the last
decade, and the improvements expected from projects
currently collecting data or in preparation.
\end{abstract}


\end{frontmatter}
\section{Introduction}
Soon after the expansion of the universe was firmly established, were
observational cosmologists already trying to detect a modification of
the expansion speed as a function of redshift.   
So confident were they that the expansion had to decelerate 
due to gravitational interaction of galaxies
that they introduced the so-called deceleration parameter
$q_0$, thought to be positive\cite{Sandage61}. 
Together with $H_0$, 
the deceleration parameter remained, for some time, 
the main cosmological parameters accessible to measurement. 
Nowadays, one prefers to describes the variation of the 
expansion of the universe in
terms of the energy density of its constituents and their equation
of state.

These first ``classical tests'' of the 
expansion involved measuring brightnesses of
galaxies, but questions concerning galaxy brightness evolution with
redshift rapidly surfaced and astronomers started looking for a better
standard candle.
The accelerated expansion was finally discovered at the very end of
the last century and came as a surprise~\cite{Riess98b,Perlmutter99}. 
The two teams of discoverers were aiming at measuring
the matter density parameter through the distance-redshift relation
of Type Ia supernovae, and faced a paradox: when they fitted a
matter-dominated cosmology to their data, the matter density parameter
had to be significantly negative.  Relying on the reproducibility of
Type Ia supernova explosion, the two projects were, by the end of the
1990, gaining access for the first time to a precise distance-redshift
relation extending to $z\sim 0.7$ (about half of the age of the universe), 
and the observed relation
favored an accelerated expansion.  This was surprising because there
is no room for an accelerated expansion in a matter-dominated
universe. However, a cosmological model mixing matter and a
cosmological constant could describe well these observations
\cite{Riess98b,Perlmutter99}. The cosmological constant enters in such
a model as a source term with static density, while the matter density
decreases with expansion.

Before the discovery of accelerated expansion, there had been
  earlier hints that matter might not constitute the dominant
  component of the universe at late times.  In 1975, assuming that the
  brightest galaxy in galaxy clusters can be used as a standard candle
  (in a way very similar to type Ia supernovae), Gunn and Tinsley
  \cite{1975Natur.257..454G} boldly suggested that the universe was
  accelerating, but also (wrongly) concluded that the total energy density
  exceeds the critical density (the density for which the universe is
  spatially flat, see \S \ref{sec:theory}), and that deuterium cannot
  be formed in the early universe.  In 1984, Peebles \cite{Peebles84}
  gathered the arguments, mostly based on matter clustering (we
  discuss the physics at \S \ref{sec:growth-theory}), in favour of a
  low matter density universe, and explicitly considered what has
  become our standard cosmological model. In 1990, the data from the
  new APM galaxy survey provided stronger evidence that the matter
  density is at most 1/3 of the critical density (see
  \cite{Maddox90,Efstathiou90} and \S \ref{sec:BAO}). And, in 1993, the
  measurement of the baryon fraction in galaxy clusters (i.e. the
  ratio of visible to total mass, see \S \ref{sec:clusters}),
  associated to the baryon density from big bang
  nucleosynthesis\footnote{Big-Bang Nucleosynthesis (BBN) refers to
    the synthesis of nuclei via the fusion of light elements in the 
    first minutes of
    the universe (see e.g. , \S 4 of \cite{KolbTurnerBook} \&
    \cite{Wagoner73}).  The primordial abundance of light elements
    (and in particular He) depend on the baryon-to-photon density
    ratio then.}
  also challenged the matter-dominated flat universe
  model \cite{White93}, favouring as well about 1/3 of the critical
  density in matter.  These indications favouring a low matter density
  universe call for some other content, when associated to the
  theoretical prejudice of a flat (hence at critical density)
  universe\footnote{Flatness is an inevitable consequence
    of the inflation theory, meant to solve some observation-based
    puzzles of hot big bang cosmology (see e.g. \S 8 of
    \cite{KolbTurnerBook} and references therein).}. In this context,
  a low matter density calls for some sort of ``complement'', although
  not necessarily causing acceleration. Note that observational evidence
  in favour of a critical matter-dominated universe was also produced
  concurrently (e.g. \cite{Loh86,NusserDekel93}). So, around 1997,
  observational cosmologists were mostly considering two possible models for the
  late-time universe: a low matter density (sub-critical) universe,
  and a critical matter-dominated universe. Our current paradigm is
  both low matter density and flat. The discovery of accelerated
  expansion then reconciled the measurements of matter density with the
  theoretical inclination for a critical universe.

The accelerated expansion raises deep issues with likely 
connections to general relativity and particle physics.  
In the framework of general relativity a fluid with a static or almost
static density may cause the acceleration of the expansion.
The expression ``dark energy'' used nowadays refers to such a hypothetical fluid.

Although a cosmological constant still accurately describes all
available large scale cosmological observations, phenomenologists have
been studying a very vast range of possible models to incarnate dark
energy. These models often involve scalar fields of some nature
inspired by a range of particle physics or quantum gravity theories.
We will not discuss these models here but refer to other papers of
this special issue \cite{kunz2012,martin2012,clarkson2012,derham2012}. 
The distinguishing feature of these models (or at
least of classes of model) is the way the density of dark energy
evolves with the expansion, commonly described using the ``equation of
state'' parameter $w$ relating pressure and density $p= w \rho$. Non
relativistic matter has $w=0$, while a fluid of constant density (e. g. the
cosmological constant) follows $w=-1$. We will report, in this review,
on recent constraints obtained on $w$ and discuss prospects for future
improvements.

The discovery of an accelerated expansion was initially relying only on
the distance-redshift relation of type Ia supernovae and the results
were questioned. Could there be dust in the distant universe
making distant supernovae appear dimmer? Were the supernovae
brightnesses evolving with redshift ?  But independent observational evidence
of an acceleration of the expansion grew rapidly. First, early ground based cosmic
microwave background (CMB, discussed in \S \ref{sec:CMB}) measurement pointing to a flat universe
\cite{Bernardis00} which was hard to reconcile with observed low mass
density without involving a non zero cosmological constant or
something alike; then, the detection of the baryon acoustic oscillations (BAO,
discussed in \S \ref{sec:BAO}) in the
galaxies two-point correlation function measured by the Sloan Digital
Sky Survey \cite{Eisenstein05}.  We now have strong evidence for an
accelerated expansion without invoking at all SNe~Ia (e.g. \S 4.1 in
\cite{Spergel07}, \cite{Blanchard06}), and, in a matter of a few years,
a new model of the universe has emerged, the ``concordance model''.  In
this model, the energy density content consists now of about a quarter of
matter and three quarters of dark energy, often assumed to be of
constant density, as observations indicate more and more tightly
\cite{Sullivan11}. The cosmological model where dark energy is 
assumed to be the cosmological constant, $\Lambda$ is called $\Lambda$CDM.

In this paper, we review the evidence for cosmic acceleration. In
\S \ref{sec:theory}, we briefly describe the cosmological framework and
introduce the observables for which currents constraints are reported
in \S \ref{sec:sne} to \ref{sec:age-of-the-universe}, from observations 
obtained using a number of
different techniques. An example of combined constraints on $w$ and 
$\Omega_M$ is shown in \S \ref{sec:combination}. 
In \S \ref{sec:future}, we briefly describe future
projects that will help better constraint the acceleration and
possibly shade new light on what could be the source of it.
We conclude in \S \ref{sec:conclusion}.

This review is part of a 5 paper special issue on Dark Energy with the companion papers being: 
The Phenomenological Approach to Modeling Dark Energy\cite{kunz2012}, 
Everything You always Wanted to Know about the Cosmological Constant (but Were Afraid to Ask)\cite{martin2012}, 
Establishing Homogeneity of the Universe in the Shadow of Dark Energy\cite{clarkson2012} 
and Galileons in the Sky\cite{derham2012}.

\section{Cosmic acceleration and dark energy \label{sec:theory}}

The cosmological principle states that the universe is homogeneous and
isotropic, and the Friedman-Lemaitre-Robertson-Walker (FLRW) metric
encodes this principle into its symmetries:
\[
ds^2 = dt^2-R^2(t)\left (\frac{dr^2}{1-kr^2}
              +r^2(d\theta^2+\sin^2\theta d\phi^2) \right )
\]
$R(t)$ is called the scale factor, and $k = -1$, $0$ or $1$, is the
sign of the spatial curvature\footnote{For textbooks covering these matters, 
we suggest \cite{KolbTurnerBook,Peacock-book-99}.}. Rather than $R(t)$,
one often referes to $a(t) \equiv R(t)/T_{\mathrm now}$. Objects with constant coordinates
$(r,\theta,\phi)$ are called comoving.  In the FLRW framework, it is
easy to show that photons emitted by comoving sources and detected by
comoving observers see their wavelength scale with $R(t)$:
$$
\frac{\lambda_\mathrm{reception}}{\lambda_\mathrm{emission}} = 
\frac{R(t_\mathrm{reception})}{R(t_\mathrm{emission})} \equiv 1+z
$$
where $z$ is the redshift of the (comoving) source. General relativity
postulates a relation between sources and the metric, which for the
FLRW metric are called the Friedman equations \cite{Friedmann24}:
\begin{eqnarray}
H^2(t) \equiv \left( \frac{\dot R}{R}  \right)^2 & = & 
    \frac{8 \pi G}{3} \rho - \frac{k}{R^2(t)} + \Lambda/3 \label{eq:friedmann_1}\\
\frac{\ddot R}{R} & = & - \frac{4\pi G}{3} (\rho+3p) + \Lambda/3 \label{eq:friedmann_2}
\end{eqnarray}
where $\Lambda$ is the cosmological constant, $\rho$ stands for the
energy density, and $p$ for the pressure. The second equation is often called
the Raychaudhuri equation. The energy conservation equation
\begin{equation}
\label{eq:energy-conservation}
\frac{d}{dt}(\rho R^3) = -3 p R^2 \dot{R}
\end{equation}
relates pressure to density evolution and applies also separately to
the various fluids in the universe. For non-relativistic matter, $\rho
R^3$ is constant and hence $p=0$.  A fluid with static density ($\dot
\rho = 0$) has $p = -\rho$. In both Friedman equations, $\Lambda$
could be summed into the density and pressure terms:
$\rho_\Lambda=-p_\Lambda\equiv \Lambda/8\pi G$.
Relation (\ref{eq:energy-conservation}) can be obtained by eliminating 
$\ddot R$ between Eq. \ref{eq:friedmann_1} and \ref{eq:friedmann_2}.   

Fluids can be characterised by a relation between $p$ and $\rho$. The
equation of state of each fluid $w_X$ is defined by $p_X=w_X\rho_X$,
and for a constant $w_X$, we have $\rho_X(t) \propto R(t)^{-3(1+w_X)}$. 
For matter $w=0$, while $w=-1$ for $\Lambda$, and $w=1/3$ for radiation. Given
the densities at one epoch (e.g. now) and the equations of state of
the fluids of the universe, one can solve the first Friedman equation
(Eq. \ref{eq:friedmann_1}) for $R(t)$. One can define the current critical 
density i.e. the density for which $k=0$, and the universe is flat:
$$
\rho_c = \frac{3 H_0^2}{8\pi G}
$$
where $H_0 = (\dot R/R)_{now}$ is the Hubble constant. One conveniently
expresses current densities in units of the current critical density:
$$
\Omega_M = \frac{\rho_M}{\rho_c} = \frac{8 \pi G \rho_M}{3 H_0^2}, \hspace{1cm}
   \Omega_\Lambda = \frac{\Lambda}{3 H_0^2}, \hspace{1cm} \Omega_k = - \frac{k}{R_0^2 H_0^2} 
$$ 
and for, e.g., a universe of matter and a cosmological constant, the first
   Friedman equation simplifies to
   $1=\Omega_M+\Omega_\Lambda+\Omega_k$. The quantity $H_0$ is often
introduced into expressions under the form $h \equiv H_0/100 km/s/Mpc$.
For example,  the matter physical density today is usually expressed
as $\Omega_M h^2$, and turns out to be better determined than
$\Omega_M$.

From Eq. \ref{eq:friedmann_2}, one notes that a matter-dominated ($p
\simeq 0$) universe or a radiation-dominated ($p>0$) universe sees its
expansion decelerate ($\ddot R<0$). 
More generally, once one integrates $\Lambda$ into
density and pressure, the deceleration parameter $q(z)$ 
can then be expressed as: 
\begin{equation}
q (z) \equiv -{\ddot R\over R H^2} = {1\over 2} \sum_i \Omega_i(z) \left[ 1+ 3w_i(z) \right]
\label{eq:q}
\end{equation}
where $\Omega_i(z)\equiv \rho_i(z)/\rho_{\rm crit}(z)$ is the fraction
of critical density of component $i$ at redshift $z$, and $w_i(z) \equiv p_i(z)/\rho_i(z)$, 
the equation 
of state of component $i$ at redshift $z$. 
The sign of $\ddot R$ is the one of the resulting $(\rho +
3p)$\footnote{In Newtonian gravity, since only masses source gravity, we
would find $\rho$ instead of $\rho+3p$ there.}. 
Therefore a fluid with $p <-\rho/3$ (i.e. $w<-1/3$) will cause the expansion to accelerate
(${\ddot R}>0$) when it comes to dominate; 
its pressure is negative and pressure sources gravity in general
relativity.   
By definition, such a
component will be called dark energy. Note also that a mixture of
matter and $\Lambda$ (a $w=-1$ dark energy) sees its expansion
accelerate as soon as $\rho_m<2\rho_{\Lambda}$.

Following Frieman {\it et al} \cite{FriemanRA&A}, we identify three
possible classes of explanations for the acceleration of the expansion:
\begin{enumerate}
\item a source term in Friedman equations with a negative enough
  equation of state, for which various forms have been proposed from
  the simple vacuum energy $\Lambda$ to more complicated time-variable
  scalar fields
\item Einstein equations of relativity need to be modified such as the
  acceleration is a manifestation of gravitational physics.  This
  requires a modification of geometric part of the Einstein equation
  rather than of the stress-energy part ("left side as opposed to
  right side of the equations").  For this to work the modifications
  have to apply on large scales only.
\item A third explanation involves dropping the assumption that the
  universe if spatially homogeneous on large scales. The idea is that
  non linear gravitational effects of spatial density fluctuations
  should alter the distance-redshift relation (see below) in such a
  way that it would explain its apparent departure from a dark energy
  free universe.
\end{enumerate}

It is not the purpose of this review to discuss the possible sources
of acceleration. Here, we will rather concentrate on discussing the
evidence for cosmic acceleration through constraints obtained the
values of $\Omega_i$ and $w_i$ as measured today. For an in-depth
review of the phenomenology associated with the specific case of 
a cosmological constant, we refer the reader to \cite{martin2012}.

The first constraints, which lead to the discovery of dark energy were
obtained using type Ia supernovae to measure the distance-redshift
relation

\subsection{Cosmological distances and comoving volume \label{sec:distances}}

With the Friedman equation, one can integrate the photon path equation
$ds=0$ for r(t), and compute various distances relevant to describe
cosmological observations \cite{CPT92}, as a function of redshift of
the emitter and the parameters describing the source terms in the
Friedman equation.  For a source emitting a (rest frame) power L and
with a measured energy flux L, the luminosity distance is defined by
$d_L(z) \equiv \sqrt{L/4\pi F}$.  Its expression as a function of
cosmological parameters reads \cite{CPT92}:
\begin{center}
\begin{eqnarray}
\hspace{-15mm}
d_L(z) &= & (1+z)H_0^{-1} |\Omega_k|^{-1/2}Sin \left\{  |\Omega_k|^{1/2} r(z) \right\} \label{eq:r_of_z} \\
r(z) &\equiv&  \int_0^z \frac{dz'}{H(z')} =  \int_0^z[ \Omega_M(1+z')^3+\Omega_\Lambda+\Omega_k(1+z')^2]^{-1/2} dz' \nonumber
\end{eqnarray}
\end{center}
where $Sin(x) = sin(x),x,sinh(x)$ for $k=1,0,-1$; note that the
expression is continuous in $\Omega_k=0$. 
This expression indicates that the distance-redshift relation probes
the source terms of the Friedman equation. A Taylor expansion around
$z=0$ reads $d_L(z) = z/H_0 + O(z^2)$, which shows that densities only
enter the expression beyond the first order in redshift. $H_0$ only
enters as a global factor in the distance expression, so that $H_0d_L$
only depends on redshift and reduced densities. One generalises
Eq. \ref{eq:r_of_z} to alternatives to $\Lambda$ (where dark energy
might have a time-variable density) by replacing the
$\Omega_{\Lambda}$ term by the (reduced) density of the considered
fluid. For a constant equation of state, $\Omega_\Lambda \rightarrow
\Omega_X(1+z)^{3(1+w)}$, and for a time-variable equation of state
w(z), $\Omega_\Lambda \rightarrow\Omega_X \ {\rm exp} [ 3 \int_0^z\frac{1+w(z')}{1+z'} dz']$.
If one considers epochs when radiation was important, one should add 
$\Omega_r(1+z)^4$ to the sum of densities.

Fig. \ref{fig:plotdl} displays the luminosity distance for a 
few cosmologies with varying admixtures of matter and cosmological constant,
corresponding to a range of acceleration values now. One can note that the
curve corresponding to our present $\Lambda$CDM paradigm cannot be mimicked
with matter-dominated distance-redshift relations.

The angular distance $d_A$ is defined via the apparent angular size
$\theta$ of an object of comoving physical size D : $d_A \equiv
D/\theta$. Because photons follow null-geodesics of the metric
\cite{Etherington33}, we have $d_L = (1+z)^2d_A$ and hence $d_L$ and $d_A$
convey the same cosmological information. $d_L$ measurements rely on
``standard candles'' while $d_A$ measurements rely on ``standard
rulers''. 

Comoving volumes can be used to constrain cosmological parameters from
e.g. counts of objects of known comoving densities. For convenience,
the comoving volume should be indexed by redshift :
\begin{equation}
\frac{d^2V}{dz\ d\Omega} = \frac{d_M^2(z)}{H(z)}
\label{eq:dvdz}
\end{equation}
where $d_M = (1+z) d_A$.

\subsection{Growth of structures \label{sec:growth-theory}}
The expansion of the universe is accompanied by the increase in
density contrast, essentially on all scales. This is called growth of
structures. The subject is considerably more complex than homogeneous
cosmologies and we will concentrate here on the salient features for
what follows and point the interested readers to, e.g., chapter 15 of
\cite{Peacock-book-99} (and references therein).

\begin{figure}[h]
\begin{center}
\begin{minipage}[t]{0.47\textwidth}
\includegraphics[width=\textwidth]{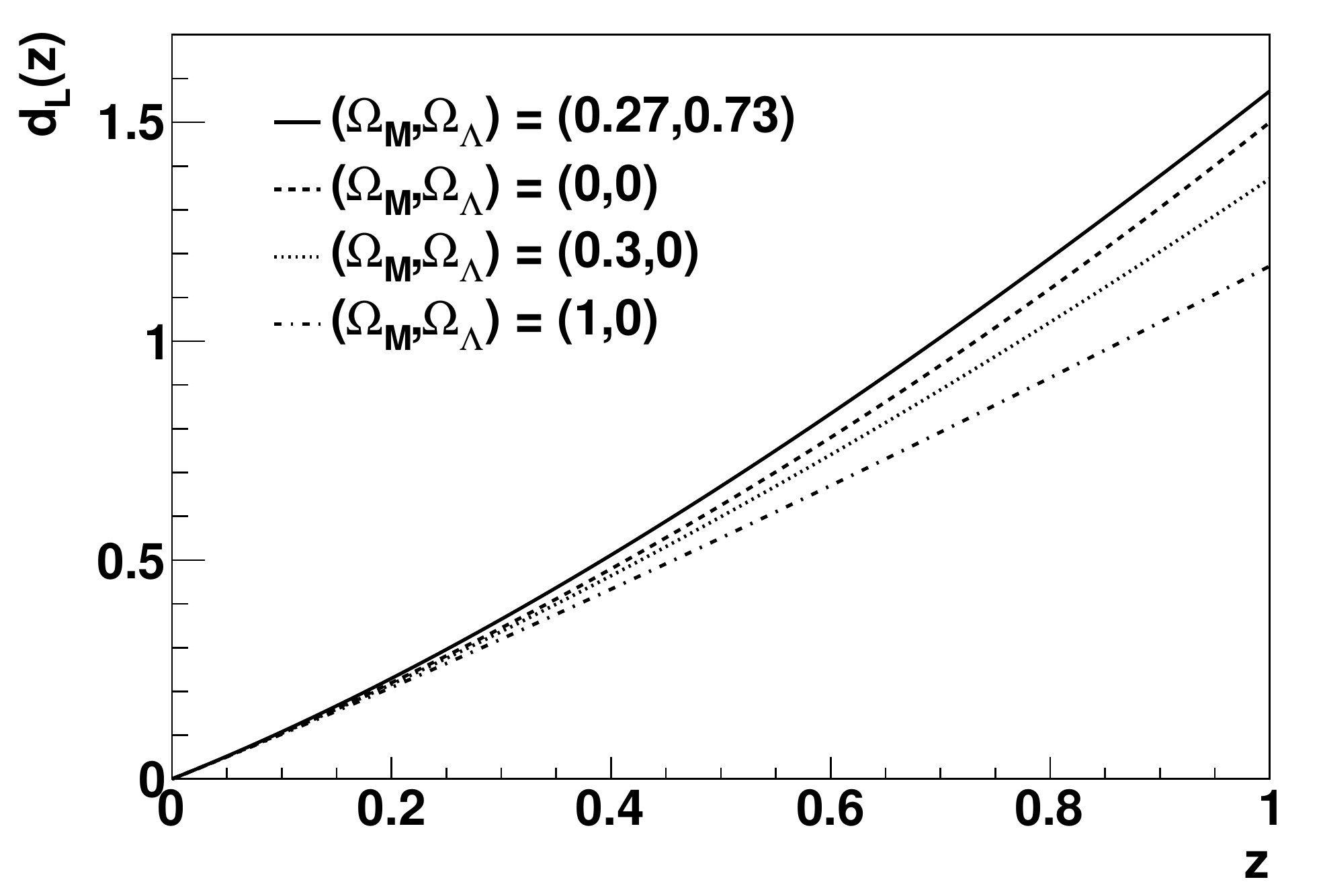}
\caption{\it Luminosity distances for a few cosmologies ranging from 
now decelerating now to now accelerating. The $\Lambda$CDM cosmology
(top curve) cannot be mimicked by any matter-dominated scenario.
\label{fig:plotdl}}
\end{minipage}\hfill
\begin{minipage}[t]{0.47\textwidth}
\includegraphics[width=\textwidth]{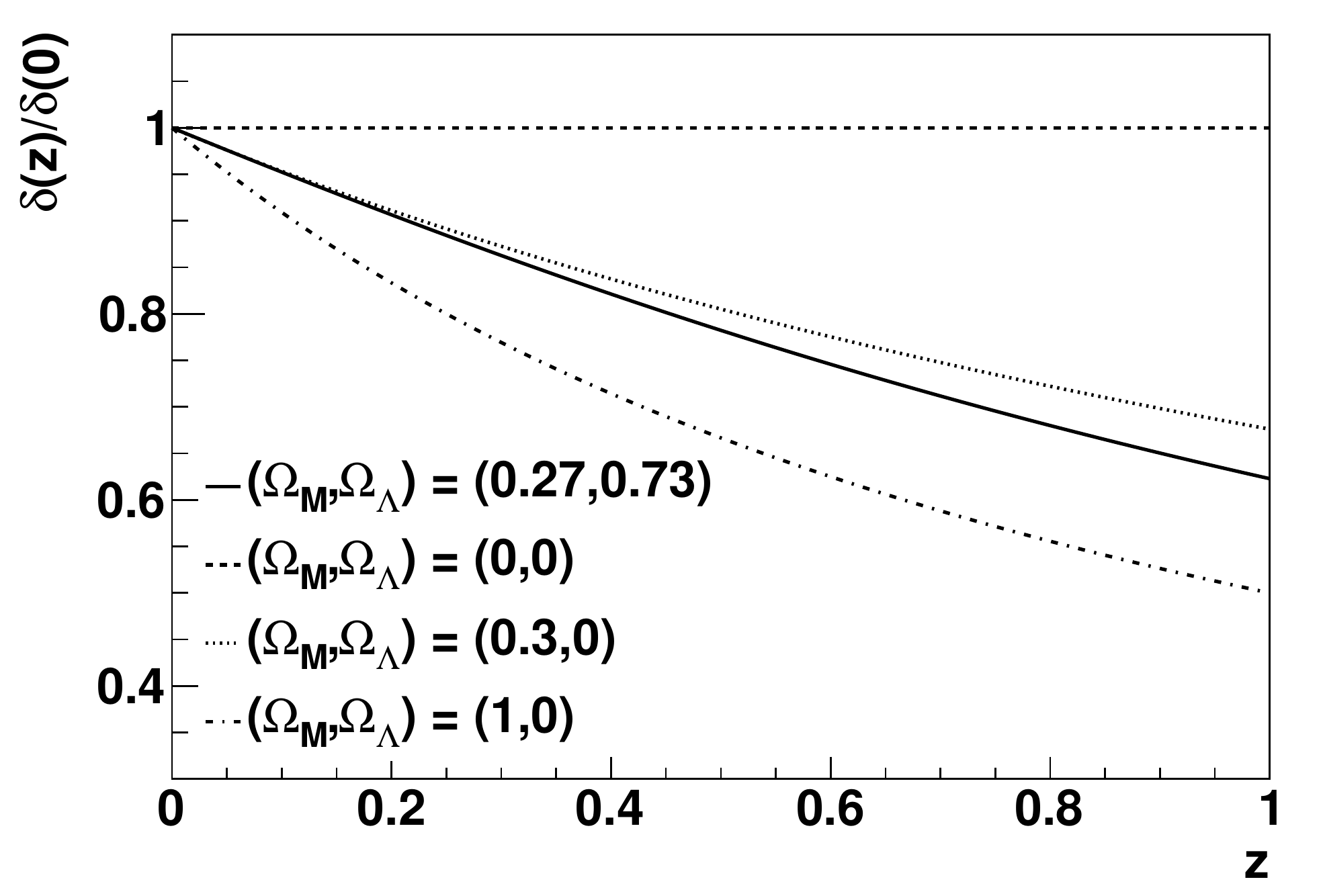}
\caption{\it Linear growth factor (growing mode) for various cosmologies. 
Note that the ordering is quite different from distances of Fig. \ref{fig:plotdl}.
\label{fig:plotgrowth}}
\end{minipage}
\end{center}
\end{figure}

Density perturbations $\delta$ are defined by $1+\delta(x) =
\rho(x)/<\rho>$.  In the late universe, matter density perturbations follow
the following differential equation:
\begin{equation}
\ddot \delta + 2 H \dot \delta = 4\pi G \rho_M \delta.
\label{eq:perturbations}
\end{equation}
where we have assumed that matter is pressure-less, radiation is
negligible and we are considering scales (well) below the Hubble
radius.  This equation is perturbative in the sense that it results
from a first order expansion, and requires in particular that
$\delta\ll 1$.  Dark energy impacts the growth of structures through
its contribution to $H$ and the evolution of $\rho_M(t)$. This equation
has two solutions, and as a rule, at most one is growing. For a
critical matter-dominated universe (an excellent approximation at
$1000>z>1$), the growing mode follows $\delta(t) \propto a(t)$, which
is called the ``linear growth of
structures''. Fig. \ref{fig:plotgrowth} displays the growth factor at late
times of the growing solution for a set of chosen cosmologies.

\begin{figure}[h]
\begin{center}
\begin{minipage}[t]{0.42\textwidth}
\includegraphics[width=\textwidth]{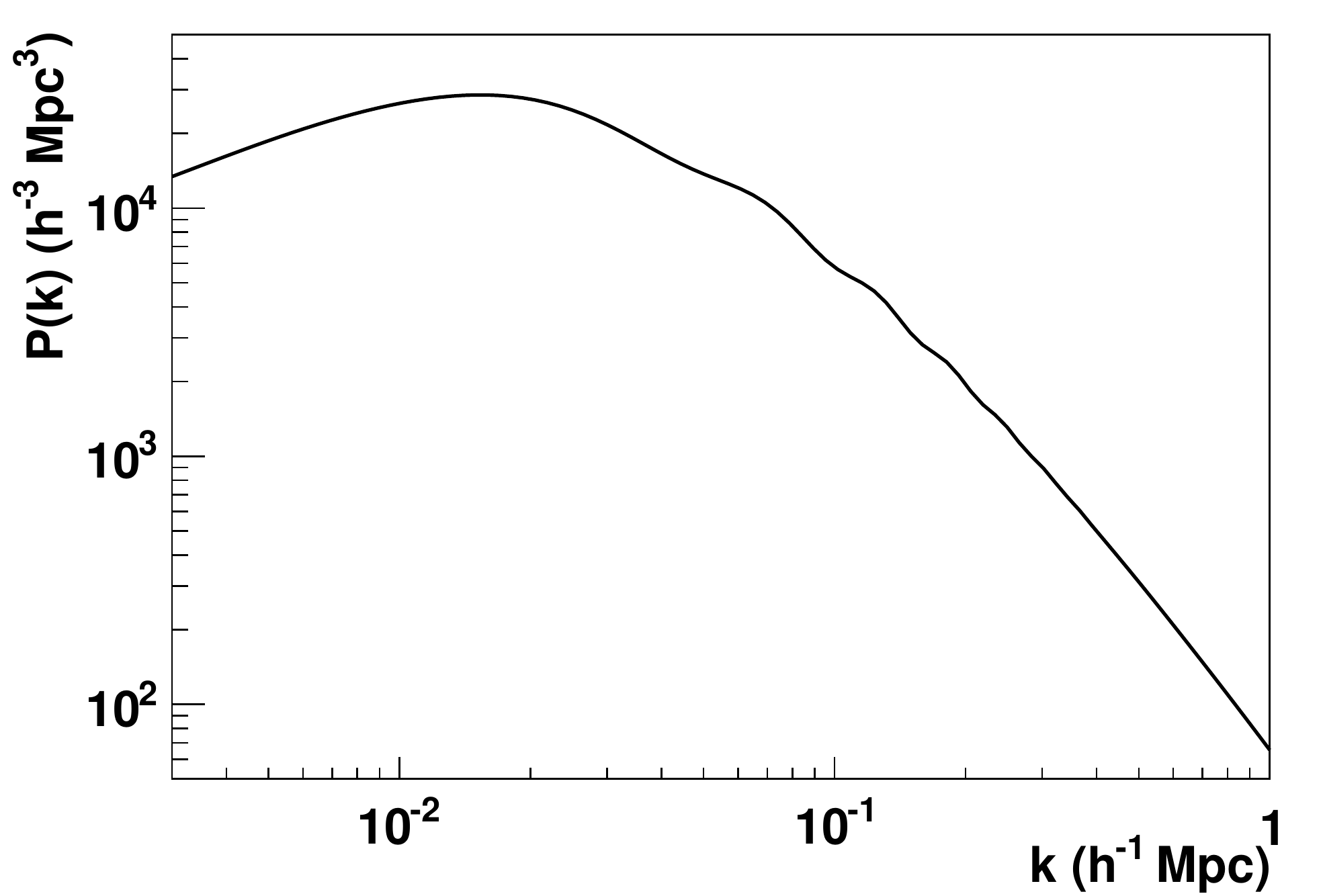}
\end{minipage}
\begin{minipage}[t]{0.49\textwidth}
\includegraphics[width=\textwidth]{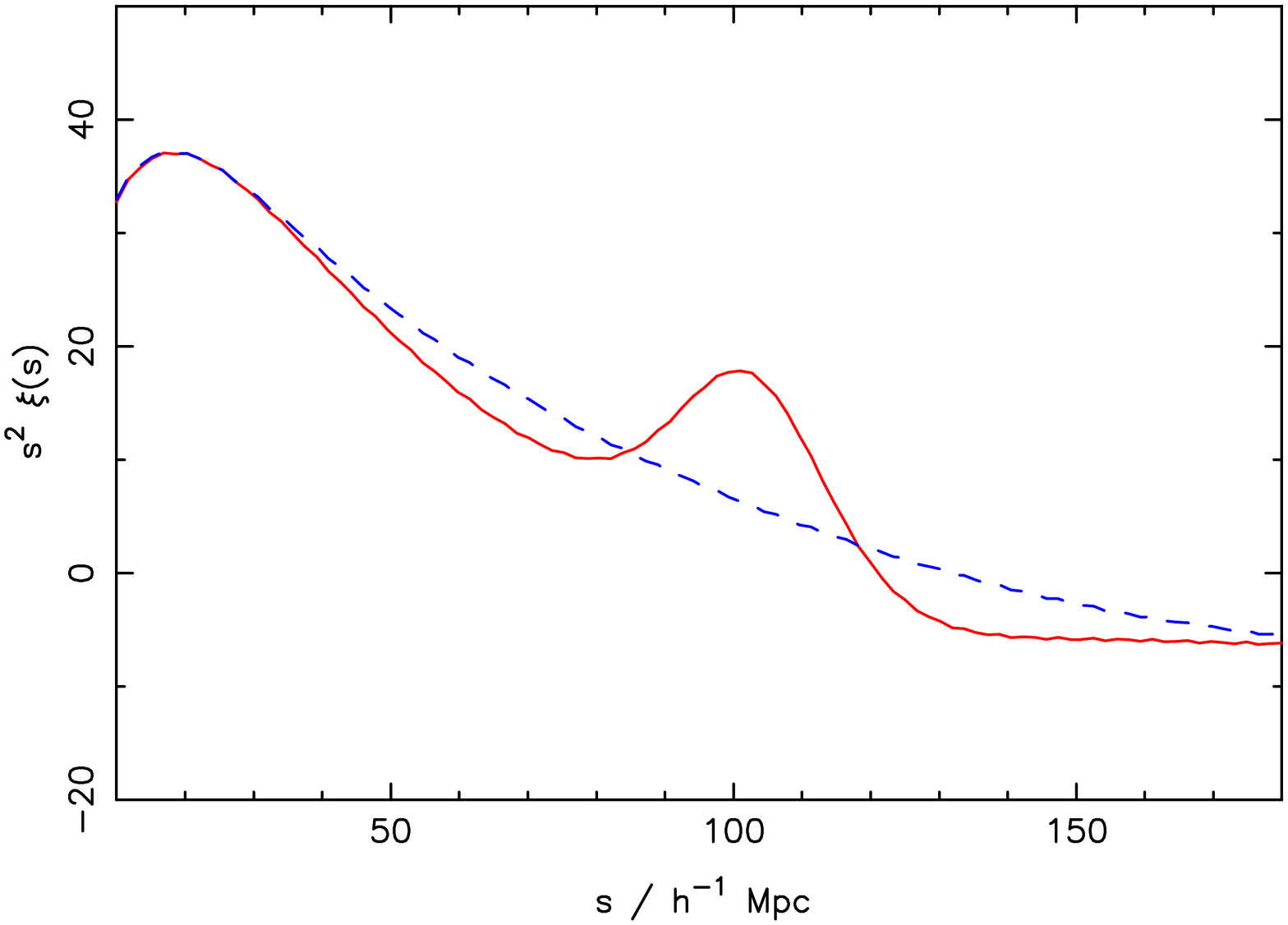}
\end{minipage}
\caption{\it 
  Left: dark matter power spectrum at z=0, computed using CMBEasy
  \cite{CMBEasy} for a flat standard $\Lambda$CDM, as a function of comoving wave number. The two key
  features are the maximum (which indicates the horizon size at
  matter-radiation equality, and depends on the matter density), and
  the series of wiggles which indicate the size of the sound horizon
  at recombination (and depends on matter and baryon densities). The
  latter corresponds to a single peak in direct space, as shown on the
  right. These features
  can be observed in Cosmic Microwave Background (CMB) anisotropies and galaxy redshift surveys,
  and their angular size constrains the expansion history. 
  Right : the matter correlation function (times $s^2$) as a function of comoving separation $s$,
  for a universe without baryons (dashed line), or with baryons (full line). 
 The acoustic peak (causing the wiggles in the power spectrum) is clearly visible (figure adapted from \cite{Blake11BAO}).\label{fig:plotps}
}
\end{center}
\end{figure}

In the early universe, matter perturbations still
follow an equation similar to Eq. \ref{eq:perturbations}
with however two differences : the expansion rate is faster
in a radiation-dominated universe, and when applicable, one should
consider the couplings of radiation to charged matter, which allow
sound waves to propagate in the primordial plasma. In a radiation-dominated
universe, matter density perturbations (both for charged and collision-free
matter) grow logarithmically with time (i.e. very slowly), and hence,
the horizon size, when matter and radiation
densities are equal (the equality epoch), is imprinted in the matter
power spectrum\footnote{Note that the electromagnetic radiation
  density is precisely known from FIRAS (aboard COBE) measurements
  \cite{Mather94} of Cosmic Microwave Background (CMB) temperature :
  $T_0=2.275 \pm 0.001 K$ \cite{Fixsen02}, and hence the CMB radiation
  density is known to 1.5 $10^{-3}$. This uncertainty has a negligible
  impact on our cosmological model \cite{Hamman08}. Since
    neutrinos contribute to radiation density at equality, their
    number density relative to photons has to be assumed for the above
    arguments to hold. In particular, we {\it assume} that there are 3
    neutrino species in the universe, i.e. that only the neutrinos
    with standard interactions (such as counted at the LEP collider) exist 
    in
    the universe. CMB precision experiments now enter in a precision
    regime that might allow this hypothesis to be tested
    (e.g. \cite{Keisler11}).}.  After equality, matter
  density perturbations grow, and the charged matter perturbations
propagate as sound waves, until the temperature is small enough to
allow atoms to form, at an epoch called ``recombination''. The travel
length of these sound waves is also imprinted in the matter
correlation function as a (slight) excess at the comoving sound
horizon size at recombination (see Fig.  \ref{fig:plotps}).

So, the clustering of matter contains two distinctive features: the
horizon size at equivalence (which scales as $\rho_{M,0}^{-1}$) and
the ``sound horizon at recombination'' (sometimes called the acoustic
scale), which is a function of the matter and baryonic matter
densities.  Fig. \ref{fig:plotps} displays the canonical matter power
spectrum and correlation function (they are related by a Fourier transform)
at low redshift, where both features are visible.

\subsection{Cosmic variance \label{cosmic-variance}}
Cosmological models do not predict the actual
observed patterns, but rather their statistical properties, such as
the average correlation function, or higher order statistics.
Unavoidably, the comparison of observations with the model is limited
by the ensemble variance of the observations, whatever their quality.
This variance floor is referred to as ``cosmic variance'', to be
understood as cosmic sampling variance. For example, the theory
predicts the average angular power spectrum of cosmic microwave
background anisotropies (see \S \ref{sec:CMB}), but we only observe a
single map and cannot practically average over the observer location.
The importance of cosmic variance in a given data set for measuring fluctuations
at a given spatial scale can be appreciated in practice by evaluating
the number of cells of this scale that the data set contains. 
The cosmic variance hence goes down when going to smaller spatial
scales.  The cosmic (sampling) variance has practical consequences: it sets a 
limit beyond which measuring more accurately the fluctuations 
will not improve significantly the cosmological constraints. For example,
some surveys have mapped the three-dimensional positions of galaxies 
in some parts of the sky, up to a certain redshift, and the galaxy counts
can be used as a proxy for the matter density field. Once the 
surveyed galaxy density is large enough for the Poisson noise
to go below the cosmic variance, surveying the same volume by other
means will not improve significantly our knowledge of e.g. the 
average power spectrum of the density field. Cosmic variance sets hard limits
about the possible measurements of matter fluctuations on large scales
in the nearby universe, and on the cosmic microwave background fluctuations
on large angular scales. For the latter, the only practical approach
consists in surveying the whole sky. 

\subsection{A brief survey of dark energy probes}

Quantifying the merits of dark energy probes and predicting
future dark energy constraints has been attempted by (at least) two
working groups and the interested reader should consult their detailed
reports \cite{DETF06,ESO-ESA}. Studying dark energy consists first in
constraining its equation of state $w$, because the cosmological
constant has $w=-1$ while other implementations give, in general,
different values. This is constrained by measuring the expansion
history, in practise the distance-redshift relation or more directly
$H(z)$.  The growth rate of structures is another handle on dark
energy, because it can deliver constraints on its own.  More
interestingly, measuring both the growth rate and the expansion
history will allow us to test General Relativity on the largest
spatial scales. All present evidence for dark energy rely on general
relativity (more precisely the Friedman equations) properly describing 
gravitation on the largest spatial scales. 

The expansion history is in practice constrained through the
distance-redshift relation. Distances can be either luminosity
distances from ``standard candles'', or angular distances from
``standard rods''. Today, the best known approximation to a standard
candle is provided by type Ia supernovae, observable at redshifts
beyond 1.  For standard rods, baryon acoustic oscillations (BAO) 
provide the ``acoustic peak'' in the correlation
function of matter, which requires large volume surveys to be detected,
and we now have measurements over a few redshift bins. Standard rods
measured across the line of sight constrain the angular distance
and may directly constrain $H(z)$ when measured along the 
line of sight. 

Gravitational lensing provides a handle on the matter distribution between 
distant galaxies and us. The ``cosmic shear'' phenomenon refers
to weak lensing by large scale structures and allows one to constrain
both distances and the matter clustering on large scales 
(and hence the growth of structures).

Studying the evolution of galaxy cluster counts with redshift allows
one to constrain both the growth of structures (in a non-linear
regime), and the expansion history through the comoving volume
(Eq. \ref{eq:dvdz}).

Supernovae, baryon acoustic oscillations, weak lensing, and clusters were the
four canonical dark energy probes studied at length in
\cite{DETF06,ESO-ESA}. They remain today the main probes of 
cosmic acceleration. 
The next section is devoted to the particular
role of CMB in dark energy constraints. 
The following sections then describe the
principle behind the measurements and current achievements of each probe.


\section{The role of the cosmic microwave background (CMB) \label{sec:CMB}}


The cosmic microwave background (CMB) plays a particular role in 
constraining the acceleration of the expansion. 
Studying anisotropies (including
polarisation) of CMB has become the key handle on cosmological
parameters. The current best determination of cosmological parameters
comes from seven years of observations with the WMAP satellite
\cite{Komatsu11}, and these results should be superseded by the Planck
satellite in early 2013.

If dark energy density evolves slowly or not at all with redshift, it
is negligible in the early universe, in particular before
recombination, the epoch of CMB emission. However, dark energy impacts the angular scale of
anisotropy correlations through our distance to the CMB emission
(called last scattering surface or LSS), because dark energy
contributes to the expansion rate at late times.  Cosmological
parameters hence contribute in two ways to the geometrical aspects of
the observed anisotropies: directly in determining the detailed
correlations of CMB anisotropies, and indirectly through our distance
to LSS (see e.g. \cite{Hu05}). The physics of the acoustic waves is
driven by the matter and baryon densities. Our distance to LSS depends
on matter density, dark energy parameters, and the Hubble constant.
At constant matter and baryon densities, and constant distance to LSS,
the observed power spectrum of fluctuations is essentially unchanged
(see e.g. \cite{Bond97}). CMB anisotropies thus provide
a single constraint on dark energy and $H_0$, which can be turned into 
a constraint on dark energy alone by using either a local measurement of $H_0$,
the flatness assumption (see e.g. \cite{Riess11} and references therein), or some other cosmological constraint.
Assuming flatness and 
that dark energy is a cosmological constant, \cite{Komatsu11} obtains
$\Omega_\Lambda=0.727\pm 0.030$. CMB constraints can of course be
integrated into fits involving more general dark energy parameters, 
e.g. \cite{Sullivan11}.

The CMB ``geometrical degeneracy'' is simply due to the fact that
observations depend on a single distance: our distance to LSS.
Secondary CMB anisotropies are the ones that build up as CMB travels
to us, and hence break this degeneracy by introducing other
distances. However, secondary anisotropies are subtle
(e.g. \cite{Aghanim08}).

CMB is affected by the gravitational deflections of foreground
structures: the image we obtain is remapped with respect to an
homogeneous universe. This phenomenon has been studied in detail 
\cite{Seljak96,Stompor99,Hu00}, and turns out to be a small
correction to most of the observables. Recently a ground-based
high resolution imager reported the detection of gravitational
lensing of the CMB \cite{Das11}. By introducing intermediate distances
between us and LSS, the phenomenon breaks the geometrical degeneracy
and allowed the same team to obtain a 3.2 $\sigma$ evidence for dark energy
{\it from CMB alone} \cite{Sherwin11}.

The integrated Sachs-Wolfe effect \cite{SachsWolfe67} is due to the
time evolution of gravitational wells when photons traverse those, and
the growth of structures is sensitive to dark energy. The Sachs-Wolfe
effect only affects large scales, which are also the most affected by
cosmic variance. Following a suggestion by \cite{CrittendenTurok95},
mild evidence for dark energy was found in \cite{Scranton03} through
the correlation of CMB temperature maps and galaxy distributions.
Recent analyses \cite{Giannantonio08,Ho08} typically reach a 4
$\sigma$ detection of dark energy through this cross-correlation
technique. Note that this detection is indeed independent of lifting
the geometrical degeneracy of CMB.

Hence, the CMB {\it on its own} is of limited use when constraining
dark energy. However since CMB constitutes the best probe to constrain
our cosmological model (see e.g. \cite{Komatsu11}), its indirect
contribution to dark energy constraints is essential
(e.g. \cite{Percival10,Amanullah10,Sullivan11}). Most if not all recent 
studies forecasting dark energy constrains (see
e.g. \cite{DETF06,ESO-ESA,EuclidRedBook11}) now assume they will
eventually use ``Planck priors'' (e.g. \cite{Mukherjee08}).

\section{Hubble diagram of SNe~Ia \label{sec:sne}}
Hubble originally published a distance-velocity
diagram\cite{Hubble29}, which was the first indication of the
expansion. We now call Hubble diagrams flux-redshift relations (or
more commonly magnitude-redshift relations), for objects of similar
intrinsic luminosity. Hubble initially reported distances to
galaxies, but the chemical evolution of galaxies seems too fast to
extend the galaxy ``Hubble diagram'' to high enough redshifts
\cite{OstrikerTremaine75}. 

The Hubble diagram of type I supernovae was
proposed to measure distances \cite{Kowal68,KirshnerKwan74} and soon
after to constrain the evolution of the expansion rate \cite{Wagoner77}.
Supernovae are explosions of stars, and their taxonomy has been
refined over the last 70 years. The current classification is
detailed in e.g. \cite{DaSilva93,Filippenko97}. Type Ia supernovae
(SNe~Ia) constitute an homogeneous subclass of supernovae, believed to
be a complete thermonuclear combustion of a white dwarf reaching the
Chandrasekhar mass (1.4 $M_\odot$), or the fusion of two white dwarfs
(for a review of our current knowledge of progenitors, see
\cite{HowellReview11}). These events are homogeneous in the sense that
they show a reproducible luminosity (see e.g. \cite{Hamuy96c}), and
can hence be used to infer luminosity distances. The luminosity rises
in about 20 rest frame days and fades over months
(e.g. \cite{LeibundgutPhD,Contreras10}) in visible bands. One usually
uses the peak brightness as the primary distance indicator, which requires
to measure the luminosity as a function of time (called ``light curve'')
in several spectral bands (see below). These events are bright enough
to be measurable up to $z\simeq 1$ using ground-based 4-m class
telescopes.

The Calan-Tololo survey delivered the first set
of precise measurements of SNe~Ia in 1996 \cite{Hamuy96b}, 29 events up to
$z\sim 0.1$, which allowed the authors to get residuals to the Hubble
diagram (magnitude-redshift) of 0.17 mag r.m.s or better\footnote{Astronomical
magnitudes are a logarithmic scale used to describe relative fluxes: $m=-2.5 log_{10}(Flux/Flux_{ref})$} (i.e. relative distance uncertainties of $\sim$8\%). These distances
make use of two empirical luminosity indicators, the decline rate of
the light curves (\cite{Pskovskii84,Phillips93}), and the
color\footnote{Colors are defined in astronomy as the ratio of
  fluxes measured in two different bands, or in practise the
  difference of magnitudes in two different bands. A typical color
  indicator, widely used for distances to supernovae and elsewhere is
  B-V, where B is a (blue) filter covering [390,480]nm, and V covers
  the green region [500,600]nm.} of the event (measured e.g. at peak
luminosity). SNe~Ia are sometimes called ``standardisable candles'',
and are the best known distance indicator to date.

In order to efficiently constrain cosmology from distances, one should use as 
long a redshift lever arm as possible: this is why cosmological constraints 
obtained from supernovae make use of nearby (typically $z<0.1$) events,
which, on their own, do not bring any interesting constraint on the expansion
evolution. Those are however vital to almost all supernova cosmology 
analyses. 

\subsection{The discovery of the accelerated expansion}

At the time the Calan-Tololo was succeeding at measuring
distances, other teams were trying to discover and measure SNe~Ia at
$z\sim 0.5$, in order to probe the distance-redshift relation beyond 
the linear regime.  Finding distant supernovae required using image subtraction
techniques \cite{Hansen87,Norgaard89,Alard98}, which consist in
digitally subtracting images of the same areas of the sky taken a few
weeks apart in order to locate light excesses. These light excesses
can then be spectroscopically identified (and their redshift
measured), and eventually their light curves measured. Both methods
and instruments for this demanding program started to be available at the
beginning of the 1990's,
and by the end of the decade, the Supernova Cosmology project 
and the High-Z teams had collected large enough 
high redshift samples to start probing the cosmic acceleration. 
The two groups compared their high redshift ($z\sim 0.5$) sample to
the Calan-Tololo nearby sample ($z<0.1$) and reached the striking
conclusion that no matter-dominated universe could describe their
magnitude-redshift relation \cite{Riess98b,Perlmutter99}.  The
measured high redshift distances are too large, compared to nearby
ones, for a decelerating universe and all matter-only universes
decelerate (see Eq. \ref{eq:friedmann_2} and Fig. \ref{fig:plotdl}).
All together the two projects had gathered $\sim$50 distant events in total, and 
the shot noise affecting their photometry was significantly degrading
the distance scatter compared to nearby events. A few years later, 
a sample of
11 events measured with the Hubble space telescope (HST), and 
featuring a photometric quality comparable to nearby events confirmed
the picture~\cite{Knop03}.

Based on this success, second generation supernovae surveys were then launched, 
which aimed at constraining a constant
equation of state of dark energy to 0.1 or better. This 
was carried out in two complementary ways: measuring distances
to very high redshift supernovae using the HST, and running 
large dedicated ground-based surveys.

\subsection{Second generation supernova surveys}
Very high redshift supernovae see their light shifted 
in the near IR, which is very difficult to observe 
from the ground because of the atmosphere large
absorption bands, and emits a bright glow. Observing
from space is then essentially mandatory for SNe~Ia at $z>1$.
A large HST program was devoted to measuring 
distances to high redshift supernovae and delivered
37 events among which 18 were at $z>1$ \cite{Riess04,Riess07}.
This sample extends deep enough in redshift to find evidence
for a past deceleration era. Nowadays, the impact of this sample 
is however limited by the modest sampling of light curves 
and the photometric calibration uncertainties (discussed in e.g. 
\S 5.1.3 of \cite{Conley11}). More recently a higher quality 
sample of 10 HST $z>1$ events was published \cite{Suzuki12},
and confirms the picture.

Ground-based wide-field imagers can efficiently tackle the $z<1$ regime
by repeatedly imaging the same area of the sky, thus building
light curves of variable objects. By tailoring the exposure time
for a $z=1$ supernova, a 1 deg$^2$ image delivers about 10 useful
measurements of SN~Ia light curves. The advent of wide-field imagers
allowed observers to propose efficient second generation SN surveys
relying on this multiplex advantage, with the promise of bringing
new constraints on the equation of state of dark energy. 
Three surveys, which benefited from large observing time allocations, 
are listed here in the order of their median redshift:
\begin{itemize}
\item The SDSS SN search used the 1.4 deg$^2$ imager on the SDSS
2.5-m telescope to monitor
300 deg$^2$ in 5 bands every second night for 3 months a year during 3 years (2005-2007).
The survey delivered light curves of $\sim 500$ spectroscopically identified 
SNe~Ia to $z=0.4$.
\item The Essence project used  the 0.36 ${\rm deg^2}$ Mosaic-II imager
on the CTIO-4m to monitor in 2 bands $\sim$ 10 ${\rm deg^2}$ for 3 months 
over 5 years (2003-2008), and has measured light curves of 228 identified SN~Ia.
\item The SNLS project relied on the 1 deg$^2$ Megacam imager on the
  3.6-m Canada-France-Hawaii Telescope. It monitored 4 pointings in 4
  bands for 5 years and measured the light curves of 450 spectroscopically identified
  SN~Ia events to $z\sim1$.
\end{itemize}
The three projects acquired as many spectra as they could to
identify spectroscopically the candidates detected in the imaging
program, and were limited by the amount of spectroscopic observing
time. They however were able to increase the statistics of distant events
by more than a factor 10, and these events
have distance accuracies that compare well with nearby ones.
During the last decade, second generation nearby SN surveys 
have also been run, and we now have about 120
high quality nearby events (see Tab 2. of \cite{Conley11}),
and more to come.

\begin{figure}[h]
\begin{center}
\begin{minipage}[t]{0.47\textwidth}
\includegraphics[width=\textwidth]{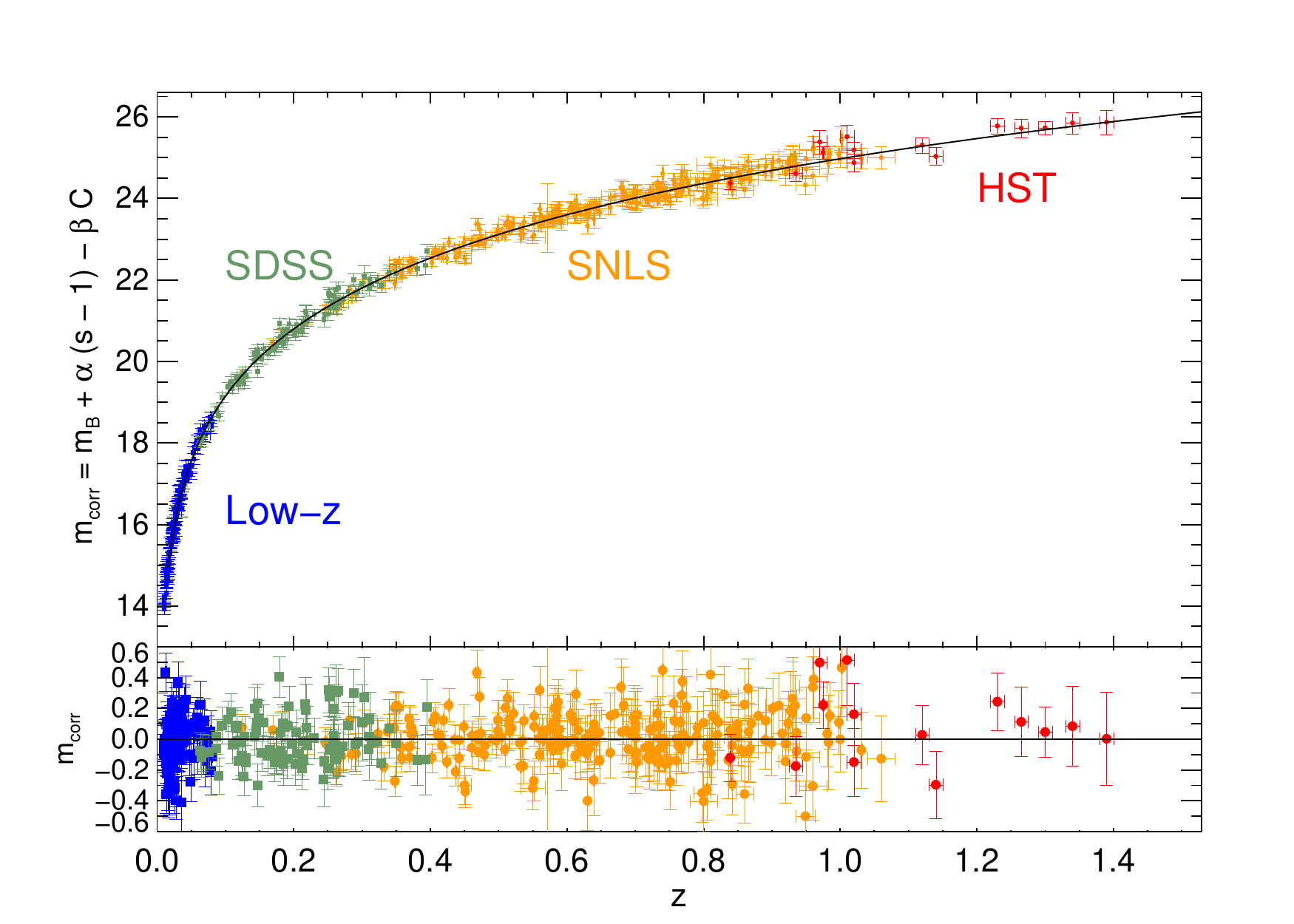}
\caption{\it Recent Hubble diagram of SNe Ia (sample assembled in
  \cite{Conley11}). The largest samples are SNLS (3 years,
  \cite{Guy10}) and SDSS (1 year, \cite{Kessler09}). The diagram
  contains about 500 events in total, and is compared to the best fit
  with supernovae only (Fig. from \cite{Sullivan11}).
\label{fig:snls3-hd}}
\end{minipage}\hfill
\begin{minipage}[t]{0.47\textwidth}
\includegraphics[width=\textwidth]{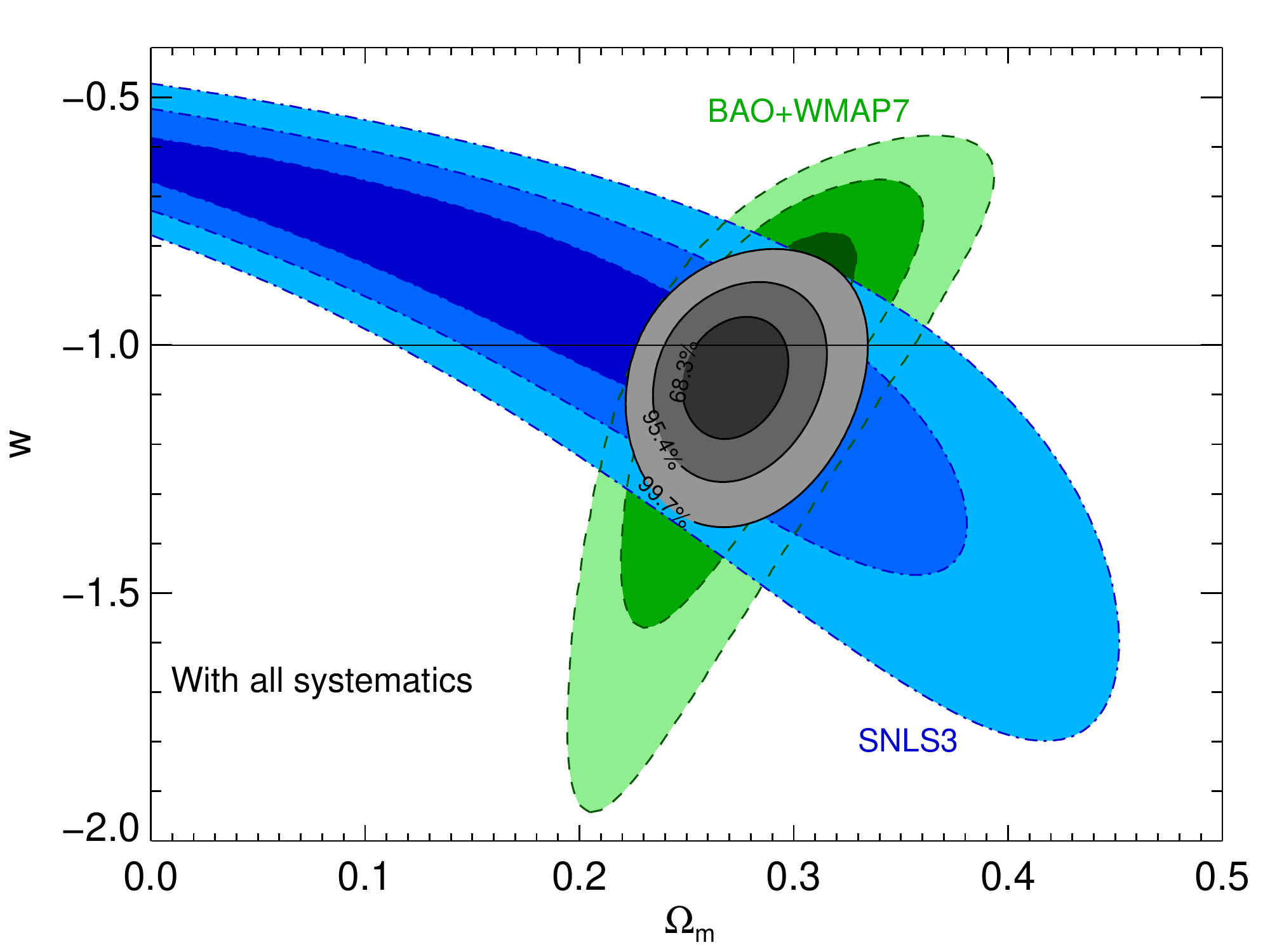}
\caption{\it Cosmological constraints obtained from the Hubble diagram of
  Fig. \ref{fig:snls3-hd} (\cite{Conley11}), from CMB
  anisotropies (\cite{Komatsu11}) and the matter power
  spectrum (\cite{Percival10}), and combined (\cite{Sullivan11}). The SN
  contours account for systematic uncertainties dominated by
  photometric calibration. Fig. from \cite{Sullivan11}.
\label{fig:snls3-w0-om}}
\end{minipage}
\end{center}
\end{figure}

So far, these 3 surveys have published partial analyses
\cite{Astier06,WoodVasey07,Kessler09,Conley11}, and none has delivered
its final sample yet. The latest compilation can be found in
\cite{Conley11} which collects 3 years of SNLS, 1 year of SDSS, and
the nearby samples, reaching a total of about 500 events passing
stringent quality cuts, including spectroscopic identification.  
The resulting Hubble diagram is shown in Fig. \ref{fig:snls3-hd}.
With a thorough accounting of systematic
uncertainties, the cosmological fit of a flat universe where
dark energy has a constant equation of state yields $w=-1.07
\pm 0.07$ \cite{Sullivan11} where the uncertainty accounts for both statistics and
systematics, which contribute almost equally. This constraint, among
the tightest to date, is illustrated on Fig. \ref{fig:snls3-w0-om}.
It is highly compatible with the cosmological constant hypothesis.

\section{Baryon acoustic oscillations (BAO) \label{sec:BAO}}

The acoustic signatures observed in the CMB anisotropies survive
recombination and leave their imprint on the matter
distribution. Namely, the correlation function of matter density
shows a peak at a comoving separation around 150 Mpc. This feature
can be used as a standard ruler and can in principle be detected both
along and across the line of sight, and yields constraints on $H(z)$
and the angular distance $d_A(z)$ respectively. Baryon acoustic
oscillations are a small signal: the probability to find a galaxy pair
at 150 Mpc separation is less than 1\% larger than at 100 or 200 Mpc;
detecting the signal require to survey at least a volume of the order
of 1 $h^{-3}Gpc^3$ (e.g. \cite{Tegmark97}).

So far, all detections made use of the galaxy distribution and merged
the longitudinal and transverse directions. The first detections were
reported in 2005 by the SDSS \cite{Eisenstein05} and the 2dF
\cite{Cole05} from the three-dimensional distribution of galaxies.
Both surveys have made use of multi-object spectrographs, which allow
one to collect hundreds of spectra at a time. Their samples of
a few 100,000 galaxies used hundreds to thousands of observing nights.
Both were redshift (i.e. 3 dimensional) galaxy surveys, limited to
$z<0.3$ for the 2dF and $0.16<z<0.47$ for the SDSS. The significance
of BAO detections is modest ($\sim$ 2.5 to 3.5 $\sigma$) but they add up
since the samples map distinct volumes. Despite this modest
significance, the SDSS measures the distance to the median redshift of
the survey ($z=0.35$) to better than 5\% using the whole correlation
function.  Multi-band imaging data from the SDSS allows one to derive
``photometric redshifts'' of galaxies, and the volume thus covered
extends to $0.2<z<0.6$ where the BAO signal is detected to 2.5
$\sigma$ \cite{Padmanabhan07}. Compared to spectroscopic redshifts,
the noise of photometric redshifts blurs the BAO peak across the line
of sight, and destroys the whole signal along the line of sight.

\begin{figure}[h]
\begin{center}
\begin{minipage}[t]{0.47\textwidth}
\includegraphics[width=\textwidth]{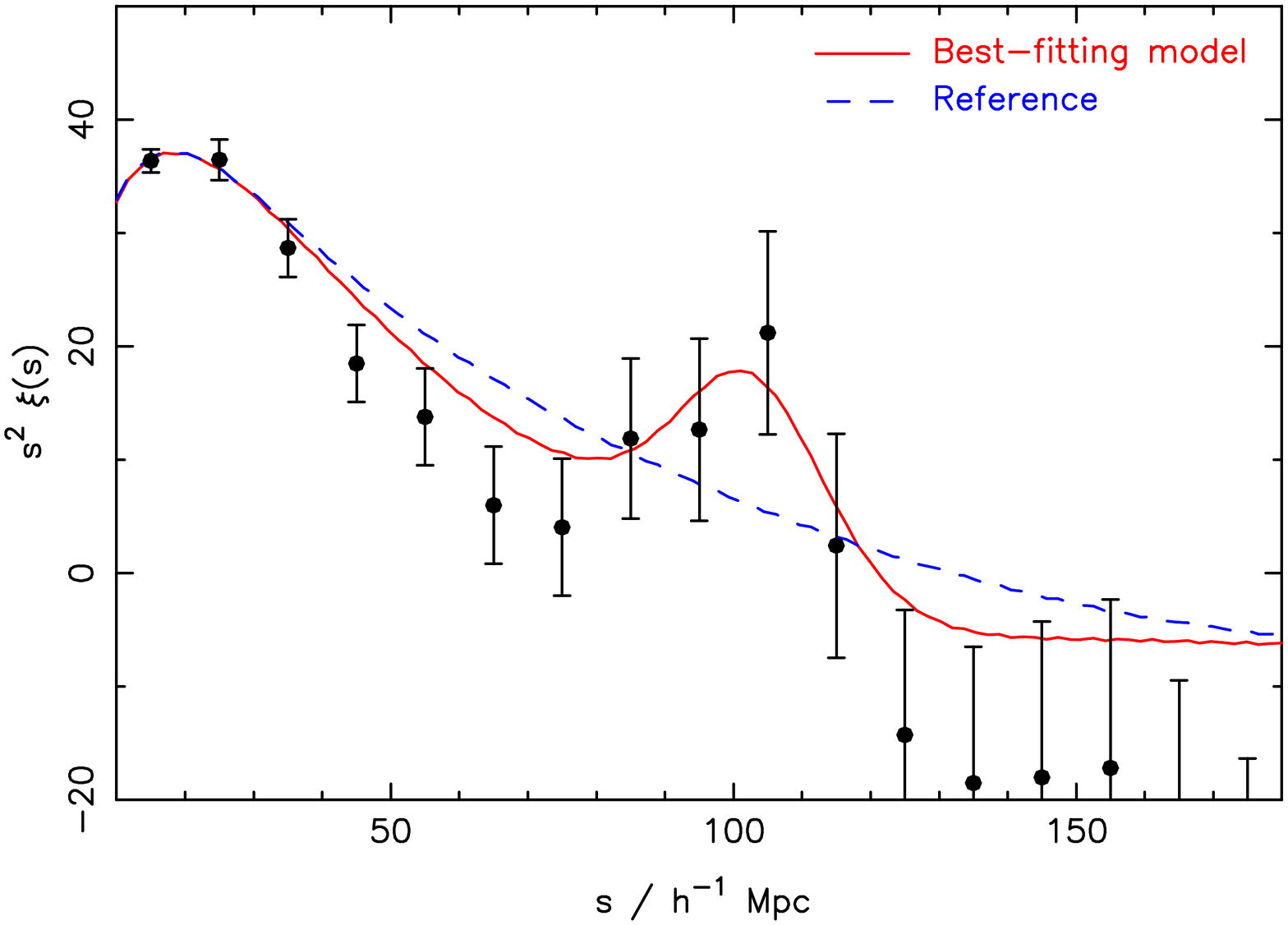}
\end{minipage}
\begin{minipage}[t]{0.47\textwidth}
\includegraphics[width=\textwidth]{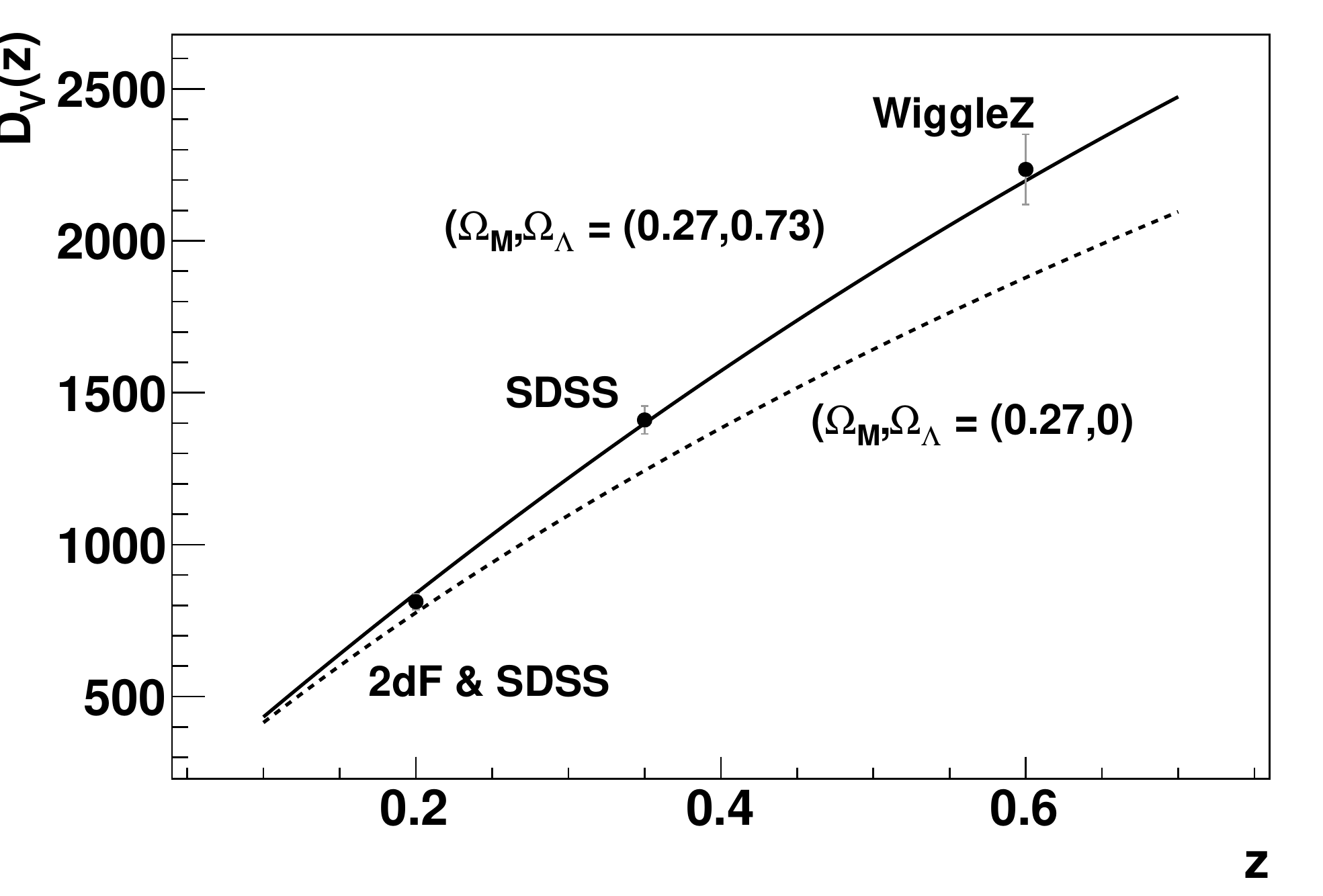}
\end{minipage}
\caption{\it 
  Right: correlation function from the WiggleZ redshift survey
(borrowed from \cite{Blake11BAO})
  as a function of comoving separation, with the
  expectations from the best fitting model, and a baryon-free
  reference model. The acoustic peak is clearly visible, but note that
  the measured points are heavily correlated.  
  Left: Distances measured in redshift slices from the acoustic
  feature in BAO surveys. The two low-redshift point come from
  \cite{Percival10} (which makes use of \cite{Cole05}), and the high
  redshift point is from \cite{Blake11BAO}. The solid curve is the
  expectation for $D_V$ (defined in Eq. \ref{eq:dv}), for a flat
  $\Lambda$CDM cosmology, where the overall scale was adjusted to the
  data. The dashed curve, does not much the data, even adjusting a 
  global factor. In \cite{Blake11-baodistances}, a similar figure is proposed
  where the WiggleZ data is split into 3 overlapping (hence
  correlated) redshift bins.
}
\label{fig:bao}
\end{center}
\end{figure}

The SDSS spectroscopic sample has been doubled since the first
detection\cite{Percival10}, and the new WiggleZ survey has published
its first results \cite{Blake11BAO}; The measured correlation
function sn shown in Fig. \ref{fig:bao}. All these
spectroscopic studies of BAO provide measurements of the acoustic
scale at various redshifts and are usually expressed using a hybrid
distance (proposed in \cite{Eisenstein05}) :
\begin{equation}
D_V = \left[ (1+z)^2 d_A^2(z) cz/H(z) \right ]^{1/3}
\label{eq:dv}
\end{equation}
which expresses that the measurement relies on two transverse 
and one longitudinal directions. The obtained constraints,
independent of the acoustic scale and the growth of structure, 
are displayed in Fig. \ref{fig:bao}:
they convey essentially the same information as the SN Hubble diagram, and
reach very compatible conclusions, but are not as precise yet.

The redshifts surveys that deliver BAO constraints can also 
efficiently constrain the matter density from the shape of the
matter power spectrum. In most of the analyses discussed above 
\cite{Cole05,Eisenstein05,Percival10,Blake11BAO}, a global fit
to the matter correlation function or power spectrum yields
essentially a constraint on $\Omega_M$, which very efficiently
complements distance measurements from SN to constrain dark energy.

The clustering of matter is sensitive to $\Omega_M$, through the
``horizon at equality'' turnover discussed in
\S~\ref{sec:growth-theory} and displayed in Fig. \ref{fig:plotps}.
Indeed, more than 20 years ago, one of the first measurements of the
angular correlation of galaxies was found to be incompatible with an
Einstein-De Sitter (flat $\Omega_M=1$) universe\cite{Maddox90,Efstathiou90},
preferring $\Omega_M \simeq 0.3$, because this delays equality
and shifts the turnover to larger spatial scales. With the prejudice of 
flatness, these results could be regarded as the first evidence 
for the presence of ``something more than just matter''.

Cosmic variance really limits the reach of BAOs at low redshift: since
the SDSS 2005 result \cite{Eisenstein05} makes use of $\sim$10\ \% of
the sky, is almost cosmic variance limited (sampling variance is about
1/3 of cosmic variance), one should not expect more than a 10 $\sigma$
whole-sky detection of BAOs at $z<\sim 0.4$, at least using similar
strategies. Similarly, cosmic variance limits the accuracy of a
distance measurement to $z\sim 0.3$ using galaxy clustering over the
whole sky to a few percent. This limitation rapidly vanishes as
redshift rises, and is totally irrelevant at $z>1$.

An interesting avenue for BAO surveys consists in accounting for the
displacement of tracers with respect to their Hubble flow positions,
due to their motions towards surrounding mass excesses. This
technique, called ``reconstruction'', was
proposed\cite{Eisenstein07}, and applied on the SDSS
data recently \cite{Padmanabhan12}, where the variance of
$D_V(z=0.35)$ is remarkably improved by about a factor of 3. This
technique is expected to be mostly effective in the low redshift
universe, where it is very welcome because cosmic variance severely
limits the ultimately achievable precision.

\section{Direct measure of the growth rate \label{sec:growth-rate-measurements}}
Three-dimensional galaxy redshift surveys allow one to probe 
the growth rate of fluctuations, assuming the expansion history is known.
One relies on the distortion of redshift from velocity
due to relative attraction of close-by galaxies which compresses their redshift 
difference\cite{Kaiser87}.
Assuming one knows $d_A(z)$ and $H(z)$,
and that galaxy clustering is isotropic on average,
one can compare the clustering across and along the line of sight
and detect these redshift distortions. For that,
one defines the nuisance bias parameter $b$, assuming the relation
$$
\left. \frac{\delta \rho}{\rho} \right|_{galaxies}  = b \left. \frac{\delta \rho}{\rho} \right|_{mass}
$$
and measurable e.g. by comparing the fluctuations of the CMB
(evolved to current epoch using Eq. \ref{eq:perturbations}) with those of
galaxy density, on spatial scales where perturbation theory holds.
One thus measures a combination of parameters $\beta \equiv f/b$
(\cite{Kaiser87}), where $f \equiv d \log \delta/d \log R$ 
describes the growth rate, and approximately reads  $\Omega_M^{0.6}$ for standard
gravity in a wide class of cosmologies around $\Lambda$CDM (e.g. p. 378 of \cite{KolbTurnerBook}).

The first evidence for redshift distortions were proposed in 2000
\cite{Hamilton00,Taylor00,Outram01}. A more precise measurement from
the two-degree-field galaxy redshift survey (2dfGRS) data (mostly at $z<0.2$)
\cite{Peacock01} (using the amplitude of CMB fluctuations to derive bias)
concludes that $\Omega_M \simeq 0.3$. Similar conclusions are reached
using a smaller sample at $z \simeq 0.55$ \cite{Ross07}.

Going to higher redshifts allows one in principle to probe the
evolution of growth rate between then and now. In \cite{Guzzo08}, the
measurement at $z\sim 0.8$ is limited by the sample size of about
10,000 galaxies of the VIPERS survey, and measures the growth rate to
2.5 $\sigma$.  On a much larger volume, the WiggleZ survey measures
the growth rate all the way to $z=0.9$ at high significance
\cite{Blake11Growth} and discusses in detail the uncertainties in the way
to account for non-linear effects, but does not venture into a fit of
cosmological parameters.

To conclude, growth rate measurements from the redshift distortions
are highly compatible with the current cosmological paradigm, not
yet at a level to significantly contribute to cosmological constrains,
and the way to overcome systematic uncertainties when measurements
get more precise is still unclear. One may note that galaxy 
redshift surveys primarily aimed at measuring BAO over very large volume and redshift intervals will deliver high quality
redshift distortion measurements in the same data sets.

\section{Clusters of galaxies \label{sec:clusters}}

The use of clusters of galaxies samples to study the acceleration of the universe 
started in the mid 1970's. At that time, brightest galaxies of clusters were used 
as standard candles 
to build Hubble diagrams extending to high enough redshifts that deviations from a 
straight Hubble line 
started to be detectable (see for example  
\cite{1975ApJ...195..255G,1978ApJ...221..383K,1976ApJ...205..688S}
).
Interestingly, Gunn and Tinsley published in 1975 a letter titled ``An accelerating Universe''
(cautiously) reporting evidence that we live in an accelerating universe      
\cite{1975Natur.257..454G}.
The conclusion was 
largely based on constraints obtained from measuring the brightness of galaxies in clusters
\cite{1975ApJ...195..255G}.

These results, however, were marginally significant and possibly subject to large systematic errors
as pointed by the authors themselves, in particular galaxy evolutionary corrections.   

Most of the constraints derived from clusters nowadays are not obtained from a fit to a Hubble diagram but rather
from the variation of the number density of clusters as a function of redshift as described below.  
 
\subsection{Clusters as cosmological probes}

In the framework of the cold dark 
matter model, the number density of dark matter halos as a function of redshift 
can be calculated and 
compared to numbers obtained in large area cluster surveys that nowadays 
extend to high enough redshift.   

Galaxy clusters are the largest virialized\footnote{In an expanding universe, strong over-densities no longer follow the expansion, and are bound. They approximately respect the virial relation between kinetic and potential energy, hence their name.} objects in the universe and are therefore expected to trace dark matter halos.
The difficulty arises from the fact that clusters are, in practice, 
selected according to some observable $O$, such as X-ray luminosity or temperature. 
Other observables often used are cluster galaxy richness, weak lensing shear or 
Sunayaev-Zeldovich effect on the cosmic microwave background flux.  
The relation of these observable $O$ selected cluster distributions with 
cluster mass distributions can be 
written as  
\begin{equation}
\frac{d^{2}N(z)}{dzd\Omega} = \frac{d_M(z)}{H(z)} 
\int^{\infty}_{0}f(O,z)dO\int^{\infty}_{0}p(O|M,z)\frac{dn(z)}{dM}dM ~,
\label{eq:clustercount}
\end{equation}
where $f(O,z)$ is the observable redshift dependent selection function, 
$dn(z)/dM$ is the comoving density of dark halos,
and $p(O|M,z)$ is the probability that a
halo of mass $M$ at redshift $z$ is observed as a cluster with observable $O$. 

Eq.~(\ref{eq:clustercount}) is sensitive to cosmology through the comoving volume element 
$ d_M^2(z)/H(z)$ (Eq \ref{eq:dvdz}) and the growth of structure term,
$dn(z)/dM$ which depends on the 
primordial spectrum and the evolution of density perturbations. 

As mentioned above, several techniques are used to detect clusters as well as for 
estimating their masses. 
Systematic uncertainties, however, can greatly affect the determination of    
the mass-observable relation $p(O|M,z)$ and of the selection function $f(O,z)$ (see for example
\cite{2009ApJ...691.1307H}).

Multi-band imaging, for example, allows clusters to be efficiently detected as excesses in the 
surface density of galaxies, and observed colors provide reliable enough redshift estimates 
that accidental projections 
can be greatly reduced. Moreover, various other effects such as weak lensing
(e.g. \cite{Mandelbaum10}), can be used to calibrate 
the mass-observable relations. Although weak lensing seems a safe approach
to evaluate cluster masses, unrelated large scale structures 
along the line of sight are a serious source of bias in this calibration,
see e.g. \cite{2002ApJ...575..640W,Hoekstra11}.

Clusters are also detected in X-ray emitted by the hot baryon gas trapped in the dark matter 
potential well, and their mass derived from X-ray luminosity or gas temperature 
(see for example
\cite{2005A&A...441..893A}).

Since it does not, in principle, depend on the source distance, the Sunayaev-Zeldovich effect 
can also be used to detect clusters out to higher redshift. 
(see for example 
\cite{2002ARA&A..40..643C}), and weak lensing can also
provide cluster detections (e.g. \cite{Miyazaki07}).

Another approach is to measure the baryonic gas mass from X-ray or SZ measurements and compare it 
with the virial mass estimates.
The ratio of the two should be independent of redshift, which can only be achieved with the correct cosmology.      

\subsection{Current cosmological constraints}
\begin{figure}[h]
\begin{center}
\begin{minipage}[t]{0.47\textwidth}
\includegraphics[width=\textwidth]{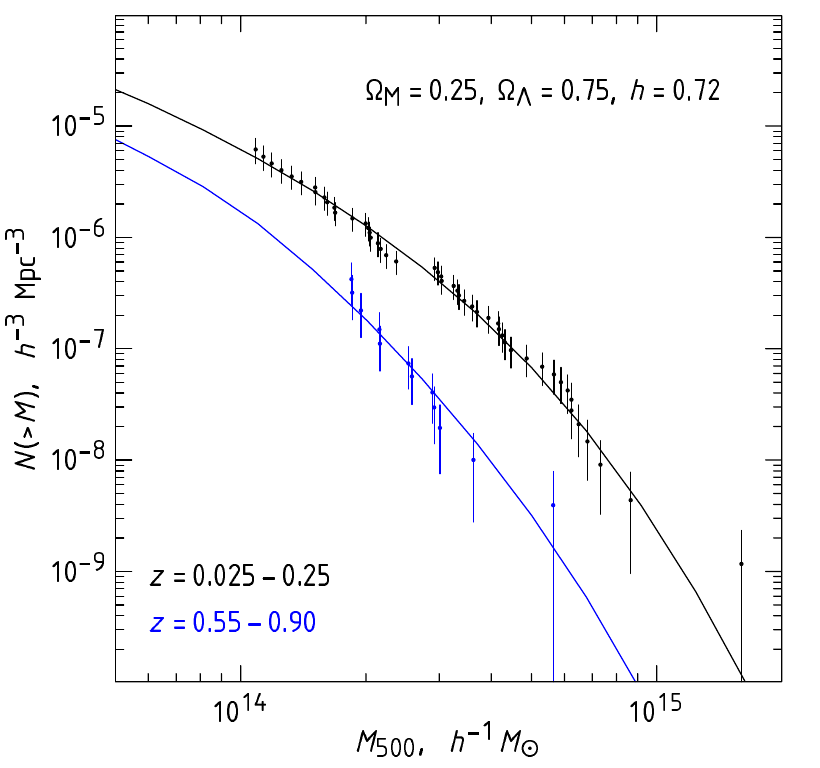}
\end{minipage}\hfill
\begin{minipage}[t]{0.47\textwidth}
\includegraphics[width=\textwidth]{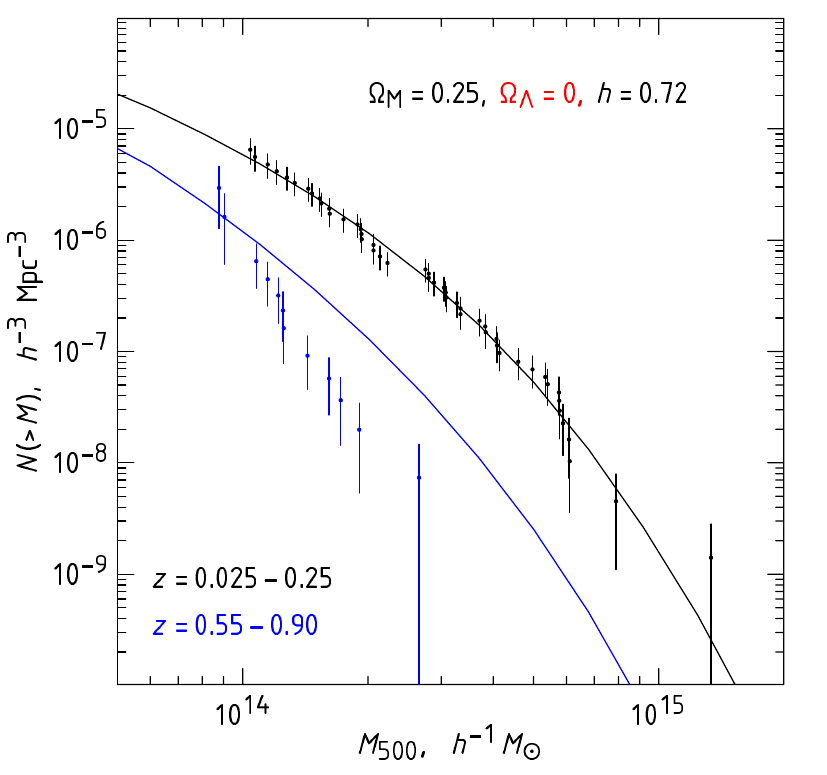}
\end{minipage}
\caption{\it 
Measured mass functions of clusters at low and high redshifts compared with 
predictions of a flat accelerating model and an open model without dark energy 
(from \cite{2009ApJ...692.1060V})
}
\label{fig:mfcn}
\end{center}
\end{figure}

Cosmological constraints obtained from clusters have greatly improved during the last decade. 
X-ray observations of clusters obtained by ``Chandra'', for example, have confirmed the acceleration
of the expansion (see \cite{2010MNRAS.406.1759M} and more recently  
\cite{2012arXiv1202.2889B}). Fig.~\ref{fig:mfcn} illustrates the power 
of cluster measurements to constraints acceleration. It shows that the measured mass 
function of clusters is correctly described provided a non zero amount of dark energy is accounted 
for in the model.     

Taken as a whole, results obtained by most of the groups and using different techniques now 
agree at the $\sim 20\%$ 
precision level on the measurements of $w$ and $\Omega_{M}$  
(see Table 2 of 
\cite{2011ARA&A..49..409A}), and are in agreement with constraints obtained with other techniques.   
Clusters are now playing an important role in constraining the cosmic acceleration.

\section{Gravitational lensing and cosmic shear \label{sec:cosmic-shear}}

Gravitational lensing refers to the deflection of light by masses, or
in a cosmological context by mass contrasts (see e.g. \S 2.4 of
\cite{Peacock-book-99}). The transverse gravitational potential
between sources and observers bends light bundles, thus remapping the image
plane, which leaves observational signatures. Strong lensing refers to
singular mappings, due to steep mass contrasts, and leads to
spectacular signatures such as giant arcs and multiple images.

Weak lensing refers to non singular mappings and has tenuous
observational signatures. The image displacement of distant galaxies
is not observable but its gradient will coherently shear their images:
locally, galaxies will display a coherent average elongation on top of
natural randomly oriented ellipticities. This ``cosmic shear'' probes
the density gradients, and the angular correlations of the cosmic
shear probe the correlations of density perturbations (see
e.g. \cite{Mellier99,BartelmannSchneider01,Refregier03Review}). The cosmic shear signal
is weak : induced ellipticities are of order 0.01, when galaxies
have natural ellipticities of $\sim$0.3 \cite{Mellier99}, and one 
has to beat down
this noise by brute force averaging. Shape distortions from the 
telescope (and atmosphere, when applicable) are commonly larger than
the signal, and foreground stars are vital to map those distortions.

Rather than the mass power spectrum itself, its evolution with redshift
(see Eq. \ref{eq:perturbations}) is sensitive to dark energy properties,
and splitting the shear signal in source redshift slices 
improves cosmological constraints \cite{Hu99}. This technique
called ``lensing tomography'', is expected to deliver
strong dark energy constraints in the future (e.g. \cite{Amara07}). For this application,
redshift of galaxies
can be approximate (however unbiased) and one uses ``photometric redshifts'' derived
from multi-band photometry (considered for this purpose as coarse spectroscopy).

First evidence of cosmic shear were found in 2000
\cite{VanWaerbeke00,Bacon00,Wittman00} from a few deg$^2$ surveys.  In
the following decade, two complementary paths were followed: improved
galaxy shape measurements from the Hubble Space Telescope (HST), and much larger surveys from
the ground.  The COSMOS field covers 1.64 deg$^2$ imaged with the HST
\cite{Scoville07}, which delivers an image quality (quantified by the size
of star images) far better than from the ground. Multi-band photometry
from UV to IR has been collected from ground and space to estimate
the photometric redshifts. Using shear tomography, a $\sim$90\% CL evidence
for acceleration was obtained from this data set \cite{Schrabback10}.

The wide CFHT legacy survey (CFHTLS, 2003-2008) has collected
images on 170 deg$^2$ in five bands from the ground. Preliminary
results do not use the tomography (\cite{Fu08} and references therein),
find a signal amplitude compatible with $\Lambda$CDM, 
measure the signal on large angular scales where perturbation
theory applies, and detect the expected signal rise with redshift of
sources.  Shear tomography results from this survey are expected 
very soon (see cfhtlens.org), but given the survey area, dark energy constraints
cannot yet outperform the current supernova and cluster counts results.

\section{Age of the universe \label{sec:age-of-the-universe}}
Given a current cosmological model, one can integrate
the Friedman equation and derive the time elapsed since the Big Bang.
The uncertainties associated to the exact nature of the ``beginning''
are totally negligible in this context. In a matter-dominated universe,
the age of the universe tends to $t_0 = H_0^{-1}$ for low matter density,
reads $t_0 = 2/3 H_0^{-1}$ for a flat universe, and has even shorter values
for closed universes. For $H_0 = 70 km/s/Mpc$, $H_0^{-1} \simeq 13$~Gy.

An observational lower limit on the age of the universe can be derived
from the confrontation of star models with real stars. The
constraint is cosmologically relevant since the oldest stars in
globular clusters have an age $12 < t_0 < 15$~Gy
\cite{KraussChaboyer03}. For a matter-dominated universe, all densities
but $\Omega_M\lesssim 0.1$ yield shorter ages, which is incompatible
with constraints from galaxy clustering (\S \ref{sec:BAO}).
With dark energy, we can
have both a $\sim$13 Gy age and a low matter density ($\Omega_M \sim 0.3$) : 
cosmological models accelerating now had a slower expansion in the past 
and hence predict a larger age than the same models without dark energy. 

The CMB anisotropies measure the distance to last scattering surface (LSS)
(i.e. when hydrogen atoms combine and light no longer scatters), which is
an increasing function of age. In a flat $\Lambda$CDM model,
the CMB anisotropies constrain $t_0 = 13.77 \pm 0.13$ \cite{Komatsu11}.
The constrain involves either the flatness assumption or some other
measurement, such as a local $H_0$ measurement.   

\section{Current constraints on dark energy \label{sec:combination}}

\begin{figure}
\begin{center}
\includegraphics[width=0.8\textwidth]{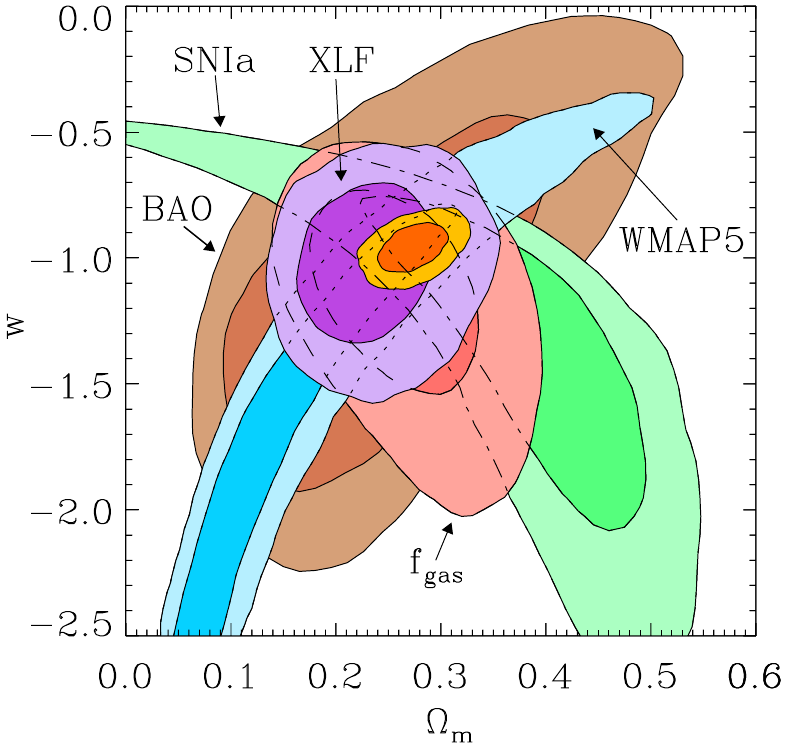}
  \caption{\it 
Joint 68.3\% and 95.4\% confidence regions for the dark energy equation of state and 
mean matter density (from \cite{2011ARA&A..49..409A})
}
  \label{fig:constraints:Mantz_w}
\end{center}
\end{figure}

Figure~\ref{fig:constraints:Mantz_w} displays a recent combination of constraints on a constant dark energy 
equation if state in a spatially flat universe (extracted from \cite{2011ARA&A..49..409A}).   
Over-plotted are constraints obtained 
from WMAP \cite{2009ApJS..180..306D}, SNIa \cite{2008ApJ...686..749K},  
BAO \cite{2010MNRAS.401.2148P}, abundance and growth of RASS clusters at $z<0.5$ 
(labeled XLF; \cite{2010MNRAS.406.1759M}) and gas fraction 
measurements at $z<1.1$ \cite{2008MNRAS.383..879A}, as well
as the combined result shown by the orange (95\%) and yellow (68\%) confidence ellipses. 
Current precisions on these parameters are at the level of 10\% both statistically and for systematics.

So far in this review, we have mainly discussed the achievements of
  observational programs in measuring the equation of state parameter
  $w$, assuming that $w$ does not vary with cosmic time.  $w$
  characterises the evolution with redshift of the dark energy
  density: for a constant $w$, $\rho_{DE} \propto (1+z)^{3(1+w)}$, and
  more generally, $\dot \rho_{DE} = -3\rho_{DE} H (1+w)$. When
  challenging the cosmological constant paradigm, or simply aiming at
  a finer characterisation of dark energy, one may consider a first
  order variation of the equation of state such as
  \cite{ChevallierPolarski01}: $w(z)\equiv w_0+w_a(1-a)=w_0+w_a
  z/(1+z)$, where $w_a$ characterises the variation. Since
  observations span at most $0<a \leq 1$, only confidence intervals
  significantly smaller than 1 are really
  constraining. Unsurprisingly, this has not happened yet, even when
  fits gather essentially all available data: \cite{Komatsu11} reports
  $w_a = -0.38 \pm 0.66$ (flat universe), and with more supernovae but
  a careful accounting of systematic uncertainties \cite{Sullivan11}
  finds $w_a = -0.984\pm 1.09$. So, the current limits on a varying
  equation of state are of limited interest. Future large projects
  such as Euclid, WFIRST or LSST could bring the uncertainty of $w_a$ down 
  to about 0.2 provided their level of systematics uncertainties are kept low.

\section{Future and prospects \label{sec:future}}

Dark energy science relies today almost entirely on imaging and spectroscopy
in the visible and near infrared.  
Several new wide-field imaging projects
are starting to take data or soon will. The Table below summarizes the
key figures of some present and future instruments and the size of 
anticipated (or executed) observing 
programs. 
Only currently approved programs are listed.
\begin{table}[h]
\begin{center}
\begin{tabular}{|l|rrrr|}
\hline
Project    & Mirror          &  Area          & First & Large survey\\
           & $\diameter$ (m) & (${\rm deg^2}$) &  light & (nights)\\
\hline
CFHT/Megacam & 3.6  & 1.    & 2002 &  500 \\
Pan-STARRS & 1.8  & 7.  & 2009 &  $>$1000 \\
Blanco/DEC & 4.0  & 3.    & 2012 & 500 \\
Subaru/HSC & 8.2 & 1.8 & 2013 &  $\sim$500 \\
LSST      & 6.5  & 10.  & 2019 & 3500 \\
Euclid    & 1.2  & 0.5  & 2019 & 5 years\\
\hline
\end{tabular}
\caption{\it 
  Key figures for the major past and future wide-field imaging
  facilities. SNLS was part of the CFHTLS, a cosmology-oriented
  500-night survey, executed at CFHT. Pan-STARRS, is currently constructing a
  second telescope and aims at eventually operating 4 of them. The
  500-night survey to be run at the CTIO-Blanco is called Dark Energy
  Survey (DES). LSST main mirror diameter is 8.4~m but suffers from a 5~m
  central occultation. This facility will almost entirely observe in
  survey mode during its anticipated 10-year lifetime. Euclid is an
  approved ESA space project meant to observe in visible and near IR.
\label{tab:wide-field-projects} 
}
\end{center}
\end{table}
They are imaging programs at the exception of the Euclid project, 
which will
carry out both imaging and spectroscopy for high redshift galaxy
survey. There are very few wide-field spectroscopy projects and 
Euclid's galaxy redshift survey is the main approved program in this field.

Regarding dark energy constraints these projects
might deliver in the future, the main forecasts
\cite{DETF06,ESO-ESA} show that systematic uncertainties
are at play for all dark energy probes, and insist
on a multi-probe approach. One should also note that
all these telescopes will deliver data sets of major
interest besides dark energy.

Second generation supernova surveys are expected to deliver their
final samples in the next two years. This will amount to about 1000
distant events (at $z>0.1$) and a growing set of nearby supernovae. 
Pan-STARSS is running a supernova program, DES is expected
to\cite{Bernstein11}, as well as LSST. Euclid does not currently have a
supernovae program, although such a program could significantly 
contribute to dark
energy constraints \cite{Astier11}.
The accuracy of cosmological constrains will primarily depend on
the accuracy of the relative photometric calibration of the various
samples.

Although weak lensing has not delivered strong dark energy constraints
yet, it concentrates hopes for the future: forecasts place the
shear correlations ahead of all other dark energy probes
(e.g. \cite{DETF06,ESO-ESA,EuclidRedBook11}). However, measuring the shear
field from galaxy shapes is a difficult problem (see e.g.
\cite{Zhang11} and references therein), and relies on a very precise
knowledge of the imaging system response
(e.g. \cite{Paulin-Henriksson09} and references therein). Current
methodologies \cite{Great08results} are improving rapidly but still behind
the required measurement accuracy \cite{Amara08}. One can now distinguish two
observational complementary approaches: high signal to noise repeated
measurements from the ground (pursued by LSST), and high resolution
images from space (pursued by Euclid).  Besides the 
difficult shear measurement, the comparison with expectations is
not straightforward on small spatial scales, because of non linearity
(e.g. \cite{Sato09}) and poorly known ``baryon'' physics
\cite{Semboloni11}. The observations required for tomographic 
shear measurements also allow one to measure cosmic magnification
statistics \cite{VanWaerbeke10}, which are free of all issues
associated to shear measurements. Both approaches probe the same 
density field but with unrelated systematics, and  can hence be 
compared without cosmic variance. 

Galaxy redshift surveys have been dominated lately by WiggleZ and the
SDSS. The SDSS telescope is currently running the BOSS program, to
deliver galaxy redshifts up to $z\sim 0.7$, from which 
results are expected very soon (see \cite{BOSS12}).
Measuring BAO at large redshifts might come from the Fastsound
redshift survey on Subaru, and also from the currently observing 
BOSS Quasar survey on SDSS-III \cite{LeGoff11}. 
Beyond these existing instruments, BigBOSS
\cite{BigBossProposal11} constitutes a natural far-reaching
ground-based follow-up project, but is not approved yet. Euclid's core
program \cite{EuclidRedBook11} includes a galaxy redshift survey
(mainly at $z>\sim 0.8$) mostly for BAO and redshift distortions.

Over the next few years, the completion of the South Pole Telescope
(SPT), the Atacama Cosmology telescope (ACT) and of the Planck SZ
surveys will result in a large increase of the number of known
clusters up to redshifts $z>1$: about 1000 new clusters are expected
to be discovered. Used in combination with existing optical and X-ray
catalogs they should lead to significant improvement of our
understanding of cluster growth and therefore further help constraint
the acceleration of the expansion. However, calibration of mass
proxies is likely to remain a limitation for these surveys and will
require improvement of spatial resolution and sensitivity planned for
the next generation of surveys such as the CCAT project.  On the
optical and near-infrared front, the large number of new ground-based
surveys (see list above) will also result in a significant increase of
the number of clusters and provide needed additional data such as the
photometric redshifts and lensing data. Here again, the difficulty
will be to define improved mass proxies in the redshift range of
interest. In space, the planned Euclid and WFIRST near-infrared
missions will allow cluster studies and measurements to be extended to
higher redshifts and larger volumes to be probed.  X-ray clusters
samples detected with Chandra and XMM-Newton will help, in the near
future, improve cosmological constraints from clusters. They will pave the
way to the use of high statistics from the eROSITA X-ray telescope, which 
is expected to detect more than 50000 clusters with unprecedented purity
and completeness.
  
\section{Summary and conclusions \label{sec:conclusion}}

Thirteen years after its discovery the acceleration of the expansion
is now firmly established and the concordance model constitutes the
frame of a standard model of cosmology.  Several techniques are now
used at telescopes around the word to probe dark energy following, and
sometimes driving, the fast development of wide-field imaging and
multi-object spectroscopy, and making use of increased precision X-ray
and CMB measurements.  Of those techniques, SN and BAO are the most
developed today, closely followed by the use of galaxy clusters, which has
made considerable progress in the last few years and the emerging
use of weak shear which promises to become one of the best tools to
measure dark energy. All these techniques make use of the impressive
precision on cosmological parameters obtained by WMAP, soon to be improved by 
Planck results. One of the key
features and power of using several techniques is that these
techniques do not always probe the same domain of the cosmological
model. Supernovae are a pure geometrical test, BAO are mostly
geometrical too, while Weak Lensing and Clusters probe both geometry and
growth of structure. Put together, they not only help break
cosmological parameter degeneracies, but more importantly are subject to 
unrelated systematics. Mixing probes could also help finding out whether the
acceleration of the expansion requires changing gravity on large scale
or not.
  
To date, constraints on the dark energy equation of state  
require combining different techniques and are at the level of 10\% both
statistically and for systematics. In the future, with the coming next
generation of experiments, each technique, combined with CMB precision
measurements from the Planck satellite, will provide individual constraints
on the dark energy equation of state, and combined they should be able
to reach percent level precision. These projects may or may not see
departure from $w=-1$ but if they do or if $w$ is found to vary with
time, they would rule out a cosmological constant or vacuum energy as
the source of acceleration and open the way to new physics. Likewise,
if the values of $w$ determined from geometry or growth of structure
methods are not equal, it would point toward a modification of gravity
as the cause of acceleration. The next decade could bring important new
observational clues on the origin of the acceleration of the
expansion.

\def\gt{$>$}
\def\aj{AJ}
\def\apj{ApJ}
\def\apjs{ApJS}
\def\apjl{ApJ}
\def\araa{ARA\&A}
\def\aa{A\&A}
\def\aap{A\&A}
\def\apss{Ap\&SS}
\def\nat{Nature}
\def\mnras{MNRAS}
\def\memras{MmRAS}
\def\aapr{The Astron. and Astrop. Rev.}
\def\prd{Phys. Rev. D}
\def\sovast{Soviet Astronomy}
\def\physrep{Phys. Rep.}
\def\jcap{J. Cosm. Astropart. P.}

\def\newblock{\ }
\bibliographystyle{unsrt3}
\bibliography{biblio,biblio-reynald}

\begin{thebibliography}{100}

\bibitem{Sandage61}
{Sandage}, A.
\newblock {The Ability of the 200-INCH Telescope to Discriminate Between
  Selected World Models.}
\newblock {\em \apj}, 133:355--+, March 1961.

\bibitem{Riess98b}
{Riess}, A.~G., {Filippenko}, A.~V., {Challis}, P., {et~al.}
\newblock {Observational Evidence from Supernovae for an Accelerating Universe
  and a Cosmological Constant}.
\newblock {\em \aj}, 116:1009--1038, September 1998.

\bibitem{Perlmutter99}
{Perlmutter}, S., {Aldering}, G., {Goldhaber}, G., {et~al.}
\newblock {Measurements of Omega and Lambda from 42 High-Redshift Supernovae}.
\newblock {\em \apj}, 517:565--586, June 1999.

\bibitem{1975Natur.257..454G}
{Gunn}, J.~E. \& {Tinsley}, B.~M.
\newblock {An accelerating Universe}.
\newblock {\em \nat}, 257:454--457, October 1975.

\bibitem{Peebles84}
{Peebles}, P.~J.~E.
\newblock {Tests of cosmological models constrained by inflation}.
\newblock {\em \apj}, 284:439--444, September 1984.

\bibitem{Maddox90}
{Maddox}, S.~J., {Efstathiou}, G., {Sutherland}, W.~J., \& {Loveday}, J.
\newblock {Galaxy correlations on large scales}.
\newblock {\em \mnras}, 242:43P--47P, January 1990.

\bibitem{Efstathiou90}
{Efstathiou}, G., {Sutherland}, W.~J., \& {Maddox}, S.~J.
\newblock {The cosmological constant and cold dark matter}.
\newblock {\em \nat}, 348:705--707, December 1990.

\bibitem{KolbTurnerBook}
{Kolb}, E.~W. \& {Turner}, M.~S.
\newblock {\em {The early universe.}}
\newblock {(Front.~Phys., Addison-Wesley)}, 1990.

\bibitem{Wagoner73}
{Wagoner}, R.~V.
\newblock {Big-Bang Nucleosynthesis Revisited}.
\newblock {\em \apj}, 179:343--360, January 1973.

\bibitem{White93}
{White}, S.~D.~M., {Navarro}, J.~F., {Evrard}, A.~E., \& {Frenk}, C.~S.
\newblock {The baryon content of galaxy clusters: a challenge to cosmological
  orthodoxy}.
\newblock {\em \nat}, 366:429--433, December 1993.

\bibitem{Loh86}
{Loh}, E.~D. \& {Spillar}, E.~J.
\newblock {A measurement of the mass density of the universe}.
\newblock {\em \apjl}, 307:L1--L4, August 1986.

\bibitem{NusserDekel93}
{Nusser}, A. \& {Dekel}, A.
\newblock {Omega and the initial fluctuations from velocity and density
  fields}.
\newblock {\em \apj}, 405:437--448, March 1993.

\bibitem{kunz2012}
Kunz, M.
\newblock {The Phenomenological Approach to Modeling Dark Energy}.
\newblock 2012.

\bibitem{martin2012}
Martin, J.
\newblock {Everything You always Wanted to Know about the Cosmological Constant
  (but Were Afraid to Ask)}.
\newblock 2012.

\bibitem{clarkson2012}
Clarkson, C.
\newblock {Establishing Homogeneity of the Universe in the Shadow of Dark
  Energy}.
\newblock 2012.

\bibitem{derham2012}
de~Rham, C.
\newblock {Galileons in the Sky}.
\newblock 2012.

\bibitem{Bernardis00}
{de Bernardis}, P., {Ade}, P.~A.~R., {Bock}, J.~J., {et~al.}
\newblock {A flat Universe from high-resolution maps of the cosmic microwave
  background radiation}.
\newblock {\em \nat}, 404:955--959, April 2000.

\bibitem{Eisenstein05}
{Eisenstein}, D.~J., {Zehavi}, I., {Hogg}, D.~W., {et~al.}
\newblock {Detection of the Baryon Acoustic Peak in the Large-Scale Correlation
  Function of SDSS Luminous Red Galaxies}.
\newblock {\em astro-ph/0501171}, January 2005.

\bibitem{Spergel07}
{Spergel}, D.~N., {Bean}, R., {Dor{\'e}}, O., {et~al.}
\newblock {Three-Year Wilkinson Microwave Anisotropy Probe (WMAP) Observations:
  Implications for Cosmology}.
\newblock {\em \apjs}, 170:377--408, June 2007.

\bibitem{Blanchard06}
{Blanchard}, A., {Douspis}, M., {Rowan-Robinson}, M., \& {Sarkar}, S.
\newblock {Large-scale galaxy correlations as a test for dark energy}.
\newblock {\em \aap}, 449:925--928, April 2006.

\bibitem{Sullivan11}
{Sullivan}, M., {Guy}, J., {Conley}, A., {et~al.}
\newblock {SNLS3: Constraints on Dark Energy Combining the Supernova Legacy
  Survey Three-year Data with Other Probes}.
\newblock {\em \apj}, 737:102, August 2011.

\bibitem{Peacock-book-99}
{Peacock}, J.~A.
\newblock {\em {Cosmological Physics}}.
\newblock {(Cambridge University Press)}, 1999.

\bibitem{Friedmann24}
Friedmann, A.
\newblock Über die möglichkeit einer welt mit konstanter negativer krümmung
  des raumes.
\newblock {\em Zeitschrift für Physik A Hadrons and Nuclei}, 21:326--332,
  1924.
\newblock 10.1007/BF01328280.

\bibitem{FriemanRA&A}
{Frieman}, J.~A., {Turner}, M.~S., \& {Huterer}, D.
\newblock {Dark Energy and the Accelerating Universe}.
\newblock {\em \araa}, 46:385--432, September 2008.

\bibitem{CPT92}
{Carroll}, S.~M., {Press}, W.~H., \& {Turner}, E.~L.
\newblock {The cosmological constant}.
\newblock {\em \araa}, 30:499--542, 1992.

\bibitem{Etherington33}
{Etherington}, I.~M.~H.
\newblock {On the Definition of Distance in General Relativity.}
\newblock {\em Philosophical Magazine}, 15:761, 1933.

\bibitem{Mather94}
{Mather}, J.~C., {Cheng}, E.~S., {Cottingham}, D.~A., {et~al.}
\newblock {Measurement of the cosmic microwave background spectrum by the COBE
  FIRAS instrument}.
\newblock {\em \apj}, 420:439--444, January 1994.

\bibitem{Fixsen02}
{Fixsen}, D.~J. \& {Mather}, J.~C.
\newblock {The Spectral Results of the Far-Infrared Absolute Spectrophotometer
  Instrument on COBE}.
\newblock {\em \apj}, 581:817--822, December 2002.

\bibitem{Hamman08}
{Hamann}, J. \& {Wong}, Y.~Y.~Y.
\newblock {The effects of cosmic microwave background (CMB) temperature
  uncertainties on cosmological parameter estimation}.
\newblock {\em \jcap}, 3:25, March 2008.

\bibitem{Keisler11}
{Keisler}, R., {Reichardt}, C.~L., {Aird}, K.~A., {et~al.}
\newblock {A Measurement of the Damping Tail of the Cosmic Microwave Background
  Power Spectrum with the South Pole Telescope}.
\newblock {\em \apj}, 743:28, December 2011.

\bibitem{CMBEasy}
{Doran}, M.
\newblock {CMBEASY: an object oriented code for the cosmic microwave
  background}.
\newblock {\em \jcap}, 10:11, October 2005.

\bibitem{Blake11BAO}
{Blake}, C., {Kazin}, E.~A., {Beutler}, F., {et~al.}
\newblock {The WiggleZ Dark Energy Survey: mapping the distance-redshift
  relation with baryon acoustic oscillations}.
\newblock {\em \mnras}, pages 1598--+, October 2011.

\bibitem{DETF06}
{Albrecht}, A., {Bernstein}, G., {Cahn}, R., {et~al.}
\newblock {Report of the Dark Energy Task Force}.
\newblock {\em ArXiv Astrophysics e-prints}, September 2006.

\bibitem{ESO-ESA}
{Peacock}, J.~A., {Schneider}, P., {Efstathiou}, G., {et~al.}
\newblock {ESA-ESO Working Group on ''Fundamental Cosmology''}.
\newblock Technical report, October 2006.

\bibitem{Komatsu11}
{Komatsu}, E., {Smith}, K.~M., {Dunkley}, J., {et~al.}
\newblock {Seven-year Wilkinson Microwave Anisotropy Probe (WMAP) Observations:
  Cosmological Interpretation}.
\newblock {\em \apjs}, 192:18--+, February 2011.

\bibitem{Hu05}
{Hu}, W.
\newblock {Dark Energy Probes in Light of the CMB}.
\newblock In {S.~C.~Wolff \& T.~R.~Lauer}, editor, {\em Observing Dark Energy},
  volume 339 of {\em Astronomical Society of the Pacific Conference Series},
  page 215, August 2005.

\bibitem{Bond97}
{Bond}, J.~R., {Efstathiou}, G., \& {Tegmark}, M.
\newblock {Forecasting cosmic parameter errors from microwave background
  anisotropy experiments}.
\newblock {\em \mnras}, 291:L33--L41, November 1997.

\bibitem{Riess11}
{Riess}, A.~G., {Macri}, L., {Casertano}, S., {et~al.}
\newblock {A 3\% Solution: Determination of the Hubble Constant with the Hubble
  Space Telescope and Wide Field Camera 3}.
\newblock {\em \apj}, 730:119, April 2011.

\bibitem{Aghanim08}
{Aghanim}, N., {Majumdar}, S., \& {Silk}, J.
\newblock {Secondary anisotropies of the CMB}.
\newblock {\em Reports on Progress in Physics}, 71(6):066902, June 2008.

\bibitem{Seljak96}
{Seljak}, U.
\newblock {Gravitational Lensing Effect on Cosmic Microwave Background
  Anisotropies: A Power Spectrum Approach}.
\newblock {\em \apj}, 463:1, May 1996.

\bibitem{Stompor99}
{Stompor}, R. \& {Efstathiou}, G.
\newblock {Gravitational lensing of cosmic microwave background anisotropies
  and cosmological parameter estimation}.
\newblock {\em \mnras}, 302:735--747, February 1999.

\bibitem{Hu00}
{Hu}, W.
\newblock {Weak lensing of the CMB: A harmonic approach}.
\newblock {\em \prd}, 62(4):043007, August 2000.

\bibitem{Das11}
{Das}, S., {Sherwin}, B.~D., {Aguirre}, P., {et~al.}
\newblock {Detection of the Power Spectrum of Cosmic Microwave Background
  Lensing by the Atacama Cosmology Telescope}.
\newblock {\em Physical Review Letters}, 107(2):021301, July 2011.

\bibitem{Sherwin11}
{Sherwin}, B.~D., {Dunkley}, J., {Das}, S., {et~al.}
\newblock {Evidence for Dark Energy from the Cosmic Microwave Background Alone
  Using the Atacama Cosmology Telescope Lensing Measurements}.
\newblock {\em Physical Review Letters}, 107(2):021302, July 2011.

\bibitem{SachsWolfe67}
{Sachs}, R.~K. \& {Wolfe}, A.~M.
\newblock {Perturbations of a Cosmological Model and Angular Variations of the
  Microwave Background}.
\newblock {\em \apj}, 147:73, January 1967.

\bibitem{CrittendenTurok95}
{Crittenden}, R.~G. \& {Turok}, N.
\newblock {Doppler Peaks from Cosmic Texture}.
\newblock {\em Physical Review Letters}, 75:2642--2645, October 1995.

\bibitem{Scranton03}
{Scranton}, R., {Connolly}, A.~J., {Nichol}, R.~C., {et~al.}
\newblock {Physical Evidence for Dark Energy}.
\newblock {\em ArXiv Astrophysics e-prints}, July 2003.

\bibitem{Giannantonio08}
{Giannantonio}, T., {Scranton}, R., {Crittenden}, R.~G., {et~al.}
\newblock {Combined analysis of the integrated Sachs-Wolfe effect and
  cosmological implications}.
\newblock {\em \prd}, 77(12):123520, June 2008.

\bibitem{Ho08}
{Ho}, S., {Hirata}, C., {Padmanabhan}, N., {Seljak}, U., \& {Bahcall}, N.
\newblock {Correlation of CMB with large-scale structure. I. Integrated
  Sachs-Wolfe tomography and cosmological implications}.
\newblock {\em \prd}, 78(4):043519, August 2008.

\bibitem{Percival10}
{Percival}, W.~J., {Reid}, B.~A., {Eisenstein}, D.~J., {et~al.}
\newblock {Baryon acoustic oscillations in the Sloan Digital Sky Survey Data
  Release 7 galaxy sample}.
\newblock {\em \mnras}, 401:2148--2168, February 2010.

\bibitem{Amanullah10}
{Amanullah}, R., {Lidman}, C., {Rubin}, D., {et~al.}
\newblock {Spectra and Light Curves of Six Type Ia Supernovae at 0.511 {\lt} z
  {\lt} 1.12 and the Union2 Compilation}.
\newblock {\em ArXiv e-prints}, April 2010.

\bibitem{EuclidRedBook11}
{Laureijs}, R., {Amiaux}, J., {Arduini}, S., {et~al.}
\newblock {Euclid Definition Study Report}.
\newblock {\em ArXiv e-prints}, October 2011.

\bibitem{Mukherjee08}
{Mukherjee}, P., {Kunz}, M., {Parkinson}, D., \& {Wang}, Y.
\newblock {Planck priors for dark energy surveys}.
\newblock {\em \prd}, 78(8):083529--+, October 2008.

\bibitem{Hubble29}
{Hubble}, E.
\newblock {A Relation between Distance and Radial Velocity among Extra-Galactic
  Nebulae}.
\newblock {\em Proceedings of the National Academy of Science}, 15:168--173,
  March 1929.

\bibitem{OstrikerTremaine75}
{Ostriker}, J.~P. \& {Tremaine}, S.~D.
\newblock {Another evolutionary correction to the luminosity of giant
  galaxies}.
\newblock {\em \apjl}, 202:L113--L117, December 1975.

\bibitem{Kowal68}
{Kowal}, C.~T.
\newblock {Absolute magnitudes of supernovae.}
\newblock {\em \aj}, 73:1021--1024, December 1968.

\bibitem{KirshnerKwan74}
{Kirshner}, R.~P. \& {Kwan}, J.
\newblock {Distances to extragalactic supernovae}.
\newblock {\em \apj}, 193:27--36, October 1974.

\bibitem{Wagoner77}
{Wagoner}, R.~V.
\newblock {Determining q0 from Supernovae}.
\newblock {\em \apjl}, 214:L5+, May 1977.

\bibitem{DaSilva93}
{da Silva}, L.~A.~L.
\newblock {The classification of supernovae}.
\newblock {\em \apss}, 202:215--236, April 1993.

\bibitem{Filippenko97}
{Filippenko}, A.~V.
\newblock {Optical Spectra of Supernovae}.
\newblock {\em \araa}, 35:309--355, 1997.

\bibitem{HowellReview11}
{Howell}, D.~A.
\newblock {Type Ia supernovae as stellar endpoints and cosmological tools}.
\newblock {\em Nature Communications}, 2, June 2011.

\bibitem{Hamuy96c}
{Hamuy}, M., {Phillips}, M.~M., {Suntzeff}, N.~B., {et~al.}
\newblock {The Hubble Diagram of the Calan/Tololo Type IA Supernovae and the
  Value of HO}.
\newblock {\em \aj}, 112:2398--+, December 1996.

\bibitem{LeibundgutPhD}
{Leibundgut}, B.
\newblock {\em {Light curves of supernovae type, I.}}
\newblock PhD thesis, PhD thesis.~Univ.~Basel.137 pp.~, (1988), 1988.

\bibitem{Contreras10}
{Contreras}, C., {Hamuy}, M., {Phillips}, M.~M., {et~al.}
\newblock {The Carnegie Supernova Project: First Photometry Data Release of
  Low-Redshift Type Ia Supernovae}.
\newblock {\em \aj}, 139:519--539, February 2010.

\bibitem{Hamuy96b}
{Hamuy}, M., {Phillips}, M.~M., {Suntzeff}, N.~B., {et~al.}
\newblock {BVRI Light Curves for 29 Type IA Supernovae}.
\newblock {\em AJ}, 112:2408--+, December 1996.

\bibitem{Pskovskii84}
{Pskovskii}, Y.~P.
\newblock {Photometric classification and basic parameters of type I
  supernovae}.
\newblock {\em \sovast}, 28:658--+, December 1984.

\bibitem{Phillips93}
{Phillips}, M.~M.
\newblock {The absolute magnitudes of Type Ia supernovae}.
\newblock {\em \apjl}, 413:L105--L108, August 1993.

\bibitem{Hansen87}
{Hansen}, L., {Jorgensen}, H.~E., \& {Norgaard-Nielsen}, H.~U.
\newblock {Search for supernovae in distant clusters of galaxies}.
\newblock {\em The Messenger}, 47:46--49, March 1987.

\bibitem{Norgaard89}
{Norgaard-Nielsen}, H.~U., {Hansen}, L., {Jorgensen}, H.~E., {Aragon
  Salamanca}, A., \& {Ellis}, R.~S.
\newblock {The discovery of a type IA supernova at a redshift of 0.31}.
\newblock {\em \nat}, 339:523--525, June 1989.

\bibitem{Alard98}
{Alard}, C. \& {Lupton}, R.~H.
\newblock {A Method for Optimal Image Subtraction}.
\newblock {\em \apj}, 503:325--+, August 1998.

\bibitem{Knop03}
{Knop}, R.~A., {Aldering}, G., {Amanullah}, R., {et~al.}
\newblock {New Constraints on {$\Omega_M$}, {$\Omega_\Lambda$}, and w from an
  Independent Set of 11 High-Redshift Supernovae Observed with the Hubble Space
  Telescope}.
\newblock {\em \apj}, 598:102--137, November 2003.

\bibitem{Riess04}
{Riess}, A.~G., {Strolger}, L., {Tonry}, J., {et~al.}
\newblock {Type Ia Supernova Discoveries at z > 1 from the Hubble Space
  Telescope: Evidence for Past Deceleration and Constraints on Dark Energy
  Evolution}.
\newblock {\em \apj}, 607:665--687, June 2004.

\bibitem{Riess07}
{Riess}, A.~G., {Strolger}, L.-G., {Casertano}, S., {et~al.}
\newblock {New Hubble Space Telescope Discoveries of Type Ia Supernovae at $z
  \geq 1$: Narrowing Constraints on the Early Behavior of Dark Energy}.
\newblock {\em \apj}, 659:98--121, April 2007.

\bibitem{Conley11}
{Conley}, A., {Guy}, J., {Sullivan}, M., {et~al.}
\newblock {Supernova Constraints and Systematic Uncertainties from the First
  Three Years of the Supernova Legacy Survey}.
\newblock {\em \apjs}, 192:1--+, January 2011.

\bibitem{Suzuki12}
{Suzuki}, N., {Rubin}, D., {Lidman}, C., {et~al.}
\newblock {The Hubble Space Telescope Cluster Supernova Survey. V. Improving
  the Dark-energy Constraints above z {\gt} 1 and Building an Early-type-hosted
  Supernova Sample}.
\newblock {\em \apj}, 746:85, February 2012.

\bibitem{Guy10}
{Guy}, J., {Sullivan}, M., {Conley}, A., {et~al.}
\newblock {The Supernova Legacy Survey 3-year sample: Type Ia supernovae
  photometric distances and cosmological constraints}.
\newblock {\em \aap}, 523:A7+, November 2010.

\bibitem{Kessler09}
{Kessler}, R., {Becker}, A.~C., {Cinabro}, D., {et~al.}
\newblock {First-Year Sloan Digital Sky Survey-II Supernova Results: Hubble
  Diagram and Cosmological Parameters}.
\newblock {\em \apjs}, 185:32--84, November 2009.

\bibitem{Astier06}
{Astier}, P., {Guy}, J., {Regnault}, N., {et~al.}
\newblock {The Supernova Legacy Survey: measurement of $\Omega_{M}$,
  $\Omega_\Lambda$ and $w$ from the first year data set}.
\newblock {\em \aap}, 447:31--48, February 2006.

\bibitem{WoodVasey07}
{Wood-Vasey}, W.~M., {Miknaitis}, G., {Stubbs}, C.~W., {et~al.}
\newblock {Observational Constraints on the Nature of Dark Energy: First
  Cosmological Results from the ESSENCE Supernova Survey}.
\newblock {\em \apj}, 666:694--715, September 2007.

\bibitem{Tegmark97}
{Tegmark}, M.
\newblock {Measuring Cosmological Parameters with Galaxy Surveys}.
\newblock {\em Physical Review Letters}, 79:3806--3809, November 1997.

\bibitem{Cole05}
{Cole}, S., {Percival}, W.~J., {Peacock}, J.~A., {et~al.}
\newblock {The 2dF Galaxy Redshift Survey: power-spectrum analysis of the final
  data set and cosmological implications}.
\newblock {\em \mnras}, 362:505--534, September 2005.

\bibitem{Padmanabhan07}
{Padmanabhan}, N., {Schlegel}, D.~J., {Seljak}, U., {et~al.}
\newblock {The clustering of luminous red galaxies in the Sloan Digital Sky
  Survey imaging data}.
\newblock {\em \mnras}, 378:852--872, July 2007.

\bibitem{Blake11-baodistances}
{Blake}, C., {Kazin}, E.~A., {Beutler}, F., {et~al.}
\newblock {The WiggleZ Dark Energy Survey: mapping the distance-redshift
  relation with baryon acoustic oscillations}.
\newblock {\em \mnras}, 418:1707--1724, December 2011.

\bibitem{Eisenstein07}
{Eisenstein}, D.~J., {Seo}, H.-J., {Sirko}, E., \& {Spergel}, D.~N.
\newblock {Improving Cosmological Distance Measurements by Reconstruction of
  the Baryon Acoustic Peak}.
\newblock {\em \apj}, 664:675--679, August 2007.

\bibitem{Padmanabhan12}
{Padmanabhan}, N., {Xu}, X., {Eisenstein}, D.~J., {et~al.}
\newblock {A 2\% Distance to z=0.35 by Reconstructing Baryon Acoustic
  Oscillations - I : Methods and Application to the Sloan Digital Sky Survey}.
\newblock {\em ArXiv e-prints}, January 2012.

\bibitem{Kaiser87}
{Kaiser}, N.
\newblock {Clustering in real space and in redshift space}.
\newblock {\em \mnras}, 227:1--21, July 1987.

\bibitem{Hamilton00}
{Hamilton}, A.~J.~S., {Tegmark}, M., \& {Padmanabhan}, N.
\newblock {Linear redshift distortions and power in the IRAS Point Source
  Catalog Redshift Survey}.
\newblock {\em \mnras}, 317:L23--L27, September 2000.

\bibitem{Taylor00}
{Taylor}, A.~N., {Ballinger}, W.~E., {Heavens}, A.~F., \& {Tadros}, H.
\newblock {Application of Data Compression Methods to the Redshift-space
  distortions of the PSCz galaxy catalogue}.
\newblock {\em ArXiv Astrophysics e-prints}, July 2000.

\bibitem{Outram01}
{Outram}, P.~J., {Hoyle}, F., \& {Shanks}, T.
\newblock {The Durham/UKST Galaxy Redshift Survey - VII. Redshift-space
  distortions in the power spectrum}.
\newblock {\em \mnras}, 321:497--501, March 2001.

\bibitem{Peacock01}
{Peacock}, J.~A., {Cole}, S., {Norberg}, P., {et~al.}
\newblock {A measurement of the cosmological mass density from clustering in
  the 2dF Galaxy Redshift Survey}.
\newblock {\em \nat}, 410:169--173, March 2001.

\bibitem{Ross07}
{Ross}, N.~P., {da {\^A}ngela}, J., {Shanks}, T., {et~al.}
\newblock {The 2dF-SDSS LRG and QSO Survey: the LRG 2-point correlation
  function and redshift-space distortions}.
\newblock {\em \mnras}, 381:573--588, October 2007.

\bibitem{Guzzo08}
{Guzzo}, L., {Pierleoni}, M., {Meneux}, B., {et~al.}
\newblock {A test of the nature of cosmic acceleration using galaxy redshift
  distortions}.
\newblock {\em \nat}, 451:541--544, January 2008.

\bibitem{Blake11Growth}
{Blake}, C., {Brough}, S., {Colless}, M., {et~al.}
\newblock {The WiggleZ Dark Energy Survey: the growth rate of cosmic structure
  since redshift z=0.9}.
\newblock {\em \mnras}, 415:2876--2891, August 2011.

\bibitem{1975ApJ...195..255G}
{Gunn}, J.~E. \& {Oke}, J.~B.
\newblock {Spectrophotometry of faint cluster galaxies and the Hubble diagram -
  an approach to cosmology}.
\newblock {\em \apj}, 195:255--268, January 1975.

\bibitem{1978ApJ...221..383K}
{Kristian}, J., {Sandage}, A., \& {Westphal}, J.~A.
\newblock {The extension of the Hubble diagram. II - New redshifts and
  photometry of very distant galaxy clusters - First indication of a deviation
  of the Hubble diagram from a straight line}.
\newblock {\em \apj}, 221:383--394, April 1978.

\bibitem{1976ApJ...205..688S}
{Sandage}, A., {Kristian}, J., \& {Westphal}, J.~A.
\newblock {The extension of the Hubble diagram. I. New redshifts and BVR
  photometry of remote cluster galaxies, and an improved richness correction.}
\newblock {\em \apj}, 205:688--695, May 1976.

\bibitem{2009ApJ...691.1307H}
{Henry}, J.~P., {Evrard}, A.~E., {Hoekstra}, H., {Babul}, A., \& {Mahdavi}, A.
\newblock {The X-Ray Cluster Normalization of the Matter Power Spectrum}.
\newblock {\em \apj}, 691:1307--1321, February 2009.

\bibitem{Mandelbaum10}
{Mandelbaum}, R., {Seljak}, U., {Baldauf}, T., \& {Smith}, R.~E.
\newblock {Precision cluster mass determination from weak lensing}.
\newblock {\em \mnras}, 405:2078--2102, July 2010.

\bibitem{2002ApJ...575..640W}
{White}, M., {van Waerbeke}, L., \& {Mackey}, J.
\newblock {Completeness in Weak-Lensing Searches for Clusters}.
\newblock {\em \apj}, 575:640--649, August 2002.

\bibitem{Hoekstra11}
{Hoekstra}, H., {Hartlap}, J., {Hilbert}, S., \& {van Uitert}, E.
\newblock {Effects of distant large-scale structure on the precision of weak
  lensing mass measurements}.
\newblock {\em \mnras}, 412:2095--2103, April 2011.

\bibitem{2005A&A...441..893A}
{Arnaud}, M., {Pointecouteau}, E., \& {Pratt}, G.~W.
\newblock {The structural and scaling properties of nearby galaxy clusters. II.
  The M-T relation}.
\newblock {\em \aap}, 441:893--903, October 2005.

\bibitem{2002ARA&A..40..643C}
{Carlstrom}, J.~E., {Holder}, G.~P., \& {Reese}, E.~D.
\newblock {Cosmology with the Sunyaev-Zel'dovich Effect}.
\newblock {\em \araa}, 40:643--680, 2002.

\bibitem{Miyazaki07}
{Miyazaki}, S., {Hamana}, T., {Ellis}, R.~S., {et~al.}
\newblock {A Subaru Weak-Lensing Survey. I. Cluster Candidates and
  Spectroscopic Verification}.
\newblock {\em \apj}, 669:714--728, November 2007.

\bibitem{2009ApJ...692.1060V}
{Vikhlinin}, A., {Kravtsov}, A.~V., {Burenin}, R.~A., {et~al.}
\newblock {Chandra Cluster Cosmology Project III: Cosmological Parameter
  Constraints}.
\newblock {\em \apj}, 692:1060--1074, February 2009.

\bibitem{2010MNRAS.406.1759M}
{Mantz}, A., {Allen}, S.~W., {Rapetti}, D., \& {Ebeling}, H.
\newblock {The observed growth of massive galaxy clusters - I. Statistical
  methods and cosmological constraints}.
\newblock {\em \mnras}, 406:1759--1772, August 2010.

\bibitem{2012arXiv1202.2889B}
{Burenin}, R.~A. \& {Vikhlinin}, A.~A.
\newblock {Cosmological parameters constraints from galaxy cluster mass
  function measurements in combination with other cosmological data}.
\newblock {\em ArXiv e-prints}, February 2012.

\bibitem{2011ARA&A..49..409A}
{Allen}, S.~W., {Evrard}, A.~E., \& {Mantz}, A.~B.
\newblock {Cosmological Parameters from Observations of Galaxy Clusters}.
\newblock {\em \araa}, 49:409--470, September 2011.

\bibitem{Mellier99}
{Mellier}, Y.
\newblock {Probing the Universe with Weak Lensing}.
\newblock {\em \araa}, 37:127--189, 1999.

\bibitem{BartelmannSchneider01}
{Bartelmann}, M. \& {Schneider}, P.
\newblock {Weak gravitational lensing}.
\newblock {\em \physrep}, 340:291--472, January 2001.

\bibitem{Refregier03Review}
{Refregier}, A.
\newblock {Weak Gravitational Lensing by Large-Scale Structure}.
\newblock {\em \araa}, 41:645--668, 2003.

\bibitem{Hu99}
{Hu}, W.
\newblock {Power Spectrum Tomography with Weak Lensing}.
\newblock {\em \apjl}, 522:L21--L24, September 1999.

\bibitem{Amara07}
{Amara}, A. \& {R{\'e}fr{\'e}gier}, A.
\newblock {Optimal surveys for weak-lensing tomography}.
\newblock {\em \mnras}, 381:1018--1026, November 2007.

\bibitem{VanWaerbeke00}
{Van Waerbeke}, L., {Mellier}, Y., {Erben}, T., {et~al.}
\newblock {Detection of correlated galaxy ellipticities from CFHT data: first
  evidence for gravitational lensing by large-scale structures}.
\newblock {\em \aap}, 358:30--44, June 2000.

\bibitem{Bacon00}
{Bacon}, D.~J., {Refregier}, A.~R., \& {Ellis}, R.~S.
\newblock {Detection of weak gravitational lensing by large-scale structure}.
\newblock {\em \mnras}, 318:625--640, October 2000.

\bibitem{Wittman00}
{Wittman}, D.~M., {Tyson}, J.~A., {Kirkman}, D., {Dell'Antonio}, I., \&
  {Bernstein}, G.
\newblock {Detection of weak gravitational lensing distortions of distant
  galaxies by cosmic dark matter at large scales}.
\newblock {\em \nat}, 405:143--148, May 2000.

\bibitem{Scoville07}
{Scoville}, N., {Abraham}, R.~G., {Aussel}, H., {et~al.}
\newblock {COSMOS: Hubble Space Telescope Observations}.
\newblock {\em \apjs}, 172:38--45, September 2007.

\bibitem{Schrabback10}
{Schrabback}, T., {Hartlap}, J., {Joachimi}, B., {et~al.}
\newblock {Evidence of the accelerated expansion of the Universe from weak
  lensing tomography with COSMOS}.
\newblock {\em \aap}, 516:A63, June 2010.

\bibitem{Fu08}
{Fu}, L., {Semboloni}, E., {Hoekstra}, H., {et~al.}
\newblock {Very weak lensing in the CFHTLS wide: cosmology from cosmic shear in
  the linear regime}.
\newblock {\em \aap}, 479:9--25, February 2008.

\bibitem{KraussChaboyer03}
{Krauss}, L.~M. \& {Chaboyer}, B.
\newblock {Age Estimates of Globular Clusters in the Milky Way: Constraints on
  Cosmology}.
\newblock {\em Science}, 299:65--70, January 2003.

\bibitem{2009ApJS..180..306D}
{Dunkley}, J., {Komatsu}, E., {Nolta}, M.~R., {et~al.}
\newblock {Five-Year Wilkinson Microwave Anisotropy Probe Observations:
  Likelihoods and Parameters from the WMAP Data}.
\newblock {\em \apjs}, 180:306--329, February 2009.

\bibitem{2008ApJ...686..749K}
{Kowalski}, M., {Rubin}, D., {Aldering}, G., {et~al.}
\newblock {Improved Cosmological Constraints from New, Old, and Combined
  Supernova Data Sets}.
\newblock {\em \apj}, 686:749--778, October 2008.

\bibitem{2010MNRAS.401.2148P}
{Percival}, W.~J., {Reid}, B.~A., {Eisenstein}, D.~J., {et~al.}
\newblock {Baryon acoustic oscillations in the Sloan Digital Sky Survey Data
  Release 7 galaxy sample}.
\newblock {\em \mnras}, 401:2148--2168, February 2010.

\bibitem{2008MNRAS.383..879A}
{Allen}, S.~W., {Rapetti}, D.~A., {Schmidt}, R.~W., {et~al.}
\newblock {Improved constraints on dark energy from Chandra X-ray observations
  of the largest relaxed galaxy clusters}.
\newblock {\em \mnras}, 383:879--896, January 2008.

\bibitem{ChevallierPolarski01}
{Chevallier}, M. \& {Polarski}, D.
\newblock {Accelerating Universes with Scaling Dark Matter}.
\newblock {\em International Journal of Modern Physics D}, 10:213--223, 2001.

\bibitem{Bernstein11}
{Bernstein}, J.~P., {Kessler}, R., {Kuhlmann}, S., {et~al.}
\newblock {Supernova Simulations and Strategies For the Dark Energy Survey}.
\newblock {\em ArXiv e-prints}, November 2011.

\bibitem{Astier11}
{Astier}, P., {Guy}, J., {Pain}, R., \& {Balland}, C.
\newblock {Dark energy constraints from a space-based supernova survey}.
\newblock {\em \aap}, 525:A7, January 2011.

\bibitem{Zhang11}
{Zhang}, J. \& {Komatsu}, E.
\newblock {Cosmic shears should not be measured in conventional ways}.
\newblock {\em \mnras}, 414:1047--1058, June 2011.

\bibitem{Paulin-Henriksson09}
{Paulin-Henriksson}, S., {Refregier}, A., \& {Amara}, A.
\newblock {Optimal point spread function modeling for weak lensing: complexity
  and sparsity}.
\newblock {\em \aap}, 500:647--655, June 2009.

\bibitem{Great08results}
{Bridle}, S., {Balan}, S.~T., {Bethge}, M., {et~al.}
\newblock {Results of the GREAT08 Challenge: An image analysis competition for
  cosmological lensing}.
\newblock {\em ArXiv e-prints}, August 2009.

\bibitem{Amara08}
{Amara}, A. \& {R{\'e}fr{\'e}gier}, A.
\newblock {Systematic bias in cosmic shear: extending the Fisher matrix}.
\newblock {\em \mnras}, 391:228--236, November 2008.

\bibitem{Sato09}
{Sato}, M., {Hamana}, T., {Takahashi}, R., {et~al.}
\newblock {Simulations of Wide-Field Weak Lensing Surveys. I. Basic Statistics
  and Non-Gaussian Effects}.
\newblock {\em \apj}, 701:945--954, August 2009.

\bibitem{Semboloni11}
{Semboloni}, E., {Hoekstra}, H., {Schaye}, J., {van Daalen}, M.~P., \&
  {McCarthy}, I.~G.
\newblock {Quantifying the effect of baryon physics on weak lensing
  tomography}.
\newblock {\em \mnras}, 417:2020--2035, November 2011.

\bibitem{VanWaerbeke10}
{van Waerbeke}, L.
\newblock {Shear and magnification: cosmic complementarity}.
\newblock {\em \mnras}, 401:2093--2100, January 2010.

\bibitem{BOSS12}
{Anderson}, L., {Aubourg}, E., {Bailey}, S., {et~al.}
\newblock {The clustering of galaxies in the SDSS-III Baryon Oscillation
  Spectroscopic Survey: Baryon Acoustic Oscillations in the Data Release 9
  Spectroscopic Galaxy Sample}.
\newblock {\em ArXiv e-prints}, March 2012.

\bibitem{LeGoff11}
{Le Goff}, J.~M., {Magneville}, C., {Rollinde}, E., {et~al.}
\newblock {Simulations of BAO reconstruction with a quasar Ly-{$\alpha$}
  survey}.
\newblock {\em \aap}, 534:A135, October 2011.

\bibitem{BigBossProposal11}
{Schlegel}, D., {Abdalla}, F., {Abraham}, T., {et~al.}
\newblock {The BigBOSS Experiment}.
\newblock {\em ArXiv e-prints}, June 2011.

\end{thebibliography}


\end{document}